\documentclass[final]{siamltex}

\usepackage{euscript,amsmath,amssymb,amsfonts,graphicx}
\allowdisplaybreaks[1]

\numberwithin{equation}{section}

\usepackage{epstopdf}

\newcommand{\e}{{\rm e}}
\renewcommand{\d}{{\rm d}}
\newcommand{\pd}{\partial}

\newcommand{\Z}{{\mathbb Z}}
\newcommand{\N}{{\mathbb N}}

\newcommand{\D}{\displaystyle}
\newcommand{\mc}{\mathcal }
\newcommand{\ve}{\varepsilon}
\newcommand{\vp}{\varphi}
\newcommand{\U}{{\mathcal U}}

\newcommand{\Lo}{{\mathcal L}}
\newcommand{\Oo}{{\mathcal O}}

\newcommand{\w}{\bar{w}}

\title{Wandering bumps in stochastic neural fields}

\author{Zachary P. Kilpatrick\thanks{Department of Mathematics, University of Pittsburgh, Pittsburgh PA ({\tt zpkilpat@pitt.edu, bard@pitt.edu})} \and Bard Ermentrout$^*$} 

\begin{document}

\maketitle

\begin{abstract} We study the effects of noise on stationary pulse solutions (bumps) in spatially extended neural fields. The dynamics of a neural field is described by an integrodifferential equation whose integral term characterizes synaptic interactions between neurons in different spatial locations of the network. Translationally symmetric neural fields support a continuum of stationary bump solutions, which may be centered at any spatial location. Random fluctuations are introduced by modeling the system as a spatially extended Langevin equation whose noise term we take to be multiplicative or additive. For nonzero noise, these bumps are shown to wander about the domain in a purely diffusive way. We can approximate the effective diffusion coefficient using a small noise expansion. Upon breaking the (continuous) translation symmetry of the system using a spatially heterogeneous inputs or synapses, bumps in the stochastic neural field can become temporarily pinned to a finite number of locations in the network. In the case of spatially heterogeneous synaptic weights, as the modulation frequency of this heterogeneity increases, the effective diffusion of bumps in the network approaches that of the network with spatially homogeneous weights.

\end{abstract}



\pagestyle{myheadings}
\thispagestyle{plain}

\section{Introduction}
\label{intro}

Spatially localized patterns of persistent neural activity (bumps) are well studied phenomena thought to subserve a variety of processes in the brain \cite{wang01}. Working (short term) memory tasks are the best known examples of brain functions that may exploit the fact that bumps are localized in feature or physical space \cite{goldmanrakic95,compte00}. For example, in oculomotor delayed-response tasks, monkeys preserve knowledge of a visual cue location using prefrontal cortical neurons with elevated activity that is correspondingly tuned to the cue location for the duration of the delay \cite{fuster73,funahashi89}. There has been a great deal of discussion concerning the relative role of various classes of prefrontal cortical neurons in maintaining persistent activity \cite{goldmanrakic95}. One strongly supported claim is that slow recurrent excitation is the operant synaptic mechanism for preserving this localized activity during the retention period \cite{wang99}. 

Experimentalists have suggested that prefrontal cortical circuitry consisting of local recurrent excitation and lateral inhibition may underlie the formation of the observed tuning of neurons to particular cue locations \cite{goldmanrakic95}. Networks with such synaptic architecture have long been studied as a theoretical framework for neural pattern formation, with seminal studies of spatially extended neural fields carried out by Wilson and Cowan \cite{wilson73} and Amari \cite{amari77}. A distinct advantage of such networks is that they display bistability, where stable spatially localized bumps can coexist with a spatially homogeneous ``off" state. Another common feature of these models is that they tend to be (continuously) translationally symmetric, since they are spatially continuous dynamical system whose symmetry is preserved under reflections and arbitrary translations \cite{bressloff12}. Stationary localized bump solutions arising in these models have been used as theoretical descriptions of tuning to visual input \cite{benyishai95,bressloff02}, head direction \cite{zhang96}, and working memory \cite{camperi98}. These studies demonstrate that neural field models are a useful tool for understanding the dynamical mechanisms necessary to sustain the neural substrates of a variety of sensory and motor processes.

Since stationary bumps are an idealized description of encoding location in networks representing feature space, many neural field studies have examined more deeply how model modifications affect the dynamics of bump solutions \cite{coombes05,bressloff12}. Many studies have also probed the effects of persistent inputs on the dynamics of neural fields with feedback inhibition \cite{amari77,benyishai97,hutt03}. For sufficiently strong inhibition, networks can generate spontaneous traveling waves so activity fails to lock to stationary \cite{benyishai97,folias04,curtu04} or traveling \cite{benyishai97,folias05,kilpatrick12} inputs. This can lead to breathing instabilities where the activity pattern oscillates regularly \cite{folias04,folias05}. Axonal delays can also substantially alter the dynamics of bumps in models with lateral inhibition, leading to multibumps \cite{coombes03}, oscillatory bumps \cite{roxin05}, and anti-pulses \cite{laing06}. Multibump solutions can also be generated by introducing synaptic connectivity that is oscillatory in space \cite{laing02,laing03}. Aside from the connectivity function, the form of the firing rate function, which converts local synaptic inputs to an output firing rate, can also affect the shape and stability of stationary bumps \cite{guo05,veltz10}. Many studies of bumps have also explored the effect of auxiliary negative feedback variables like spike frequency adaptation \cite{pinto01,coombes05b} or synaptic depression \cite{york09,kilpatrick10}. Substantially strong negative feedback can lead either a drift instability, where the bump propagates as a traveling pulse \cite{laing01b,pinto01,coombes05b,york09,kilpatrick10}, or a breathing instability, where the edges of the bump oscillate their position in time \cite{pinto01b,coombes05b,coombes12b}. Recently, it was shown that an auxiliary synaptic facilitation variable can serve to curtail the tendency of bumps in neural fields with heterogeneous connectivity to wander \cite{itskov11}. Thus, there is a veritable wealth of dynamic instabilities of bumps that have been examined in deterministic neural fields.

Beyond these studies, there have been several analyses of spiking neuron models of stationary bumps \cite{camperi98,compte00,laing01}. Spiking models have the advantage of capturing finer timescale dynamics, for example spike time synchrony, than those of which neural fields are capable. Another major difference is that spiking models are often chaotic, leading to dynamics that can appear random. This is much more akin to the environment of networks of neurons in the brain, seething with fluctuations. As a result, a basic behavior that has been revealed in numerical simulations of bumps in spiking networks is wandering of the bump's mean position \cite{camperi98,compte00,laing01}. There has been very limited investigation of such dynamics in neural field equations \cite{camperi98,itskov11}. Nonetheless, in both spiking models and neural fields with noise, the variance of the bump's position scales linearly with time, suggesting the position as a function of time behaves as a purely diffusive process \cite{camperi98,compte00,renart03,chow06}. This is due in part to these system often being translationally symmetric \cite{compte00,laing01,brody03}. While this symmetry allow bumps to be initially nucleated at any point in the network, an inherent marginal stability makes it so that bumps are never firmly pinned to any particular location over time \cite{camperi98,compte00,laing01}. Thus, bump position is fragile to noise and as well as perturbations of the evolution equations of the underlying dynamical system, which itself contains a line attractor.

The fact that bumps in noisy models of working memory wander should be no surprise, in light of existing data concerning the dependence of recall error on delay time \cite{white94,ploner98}. In spite of the relatively reliable correspondence between the elevation of neural activity and the cue location in prefrontal cortical networks \cite{goldmanrakic95}, there is inevitably some error made in reporting the original cue location \cite{white94}. Interestingly, the amplitude of this error scales linearly in time \cite{ploner98}, suggesting that it may be generated by some underlying diffusive process. Thus, for a network to have improve memory storage accuracy, it should reduce the effects of this diffusion as much as possible. This invites the question of how networks for working memory may exploit dynamics that are close to line attractors to improve memory recall accuracy. Some computational studies have suggested that relaxing the translation symmetry of line attractors by introducing multiple discrete attractors may make dynamics more resilient \cite{seung96,koulakov02,brody03}. However, others have viewed spatial heterogeneity in networks as a detriment to working memory that must be overcome \cite{renart03,itskov11}. Therefore, to make the theory of bump attractors for working memory more robust, we must consider the effects of noise and network heterogeneity and any new phenomena they bring.

We propose to perform an in depth analysis of the diffusion of stationary bump solutions in neural field equations with noise. In doing so, we wish to understand how parameters of the model affect the degradation of the bump's initial position. Since oculomotor delayed-response tasks usually require recalling the location of an object on a circle, this suggests using a neural field model whose spatial domain is finite and periodic \cite{benyishai97,camperi98,bressloff02,veltz10}. Thus, to accompany our analysis of stochastic neural fields, we will review and extend some of the results for bump existence and stability in the deterministic ring model \cite{somers95,benyishai95,bressloff02}
\begin{align}
\frac{\pd u(x,t)}{\pd t} &= -u(x,t) + \int_{- \pi}^{\pi} w(x,y) f(u(y,t)) \d y, \label{ring}
\end{align}
where $u(x,t)$ is the total synaptic input to spatial location $x \in  [- \pi , \pi ]$ at time $t$. A variation of this model, that includes spatially varying inputs, is examined in section \ref{input}. The integral term represents synaptic feedback from the entirety of the network so that the kernel $w(x,y)$ encodes the strength of connections from $y$ to $x$. In many studies of the ring model, $w(x,y) = \w (x-y)$ so the network is spatially homogeneous \cite{benyishai97,camperi98,hansel98,bressloff02,veltz10,kilpatrick12}. In particular, using the cosine weight kernel
\begin{align}
w (x,y) = \w (x-y) = \cos (x-y).     \label{cos}
\end{align}
makes the equation (\ref{ring}) translationally symmetric in space and amenable to explicit analysis. We study bump solutions that arise in the case of homogeneous synaptic weights extensively in section \ref{bump1}. However, we also study the effect of spatially heterogeneous synaptic connections in section \ref{inhom}, so that $w(x,y) = (1 + \sigma w_1(ny)) \w ( x- y).$ Particularly, we analyze (\ref{ring}) when
\begin{align}
w(x,y) = (1 + \sigma \cos (n y)) \cos (x - y),   \label{cosinh}
\end{align}
which provides spatially heterogeneous, yet periodic, synaptic modulation whose frequency is set by the $n \in \N$. Spatial heterogeneities in the weight functions of neural fields have been shown to lead to multibump solutions \cite{laing02,laing03} and to alter traveling waves \cite{bressloff01,kilpatrick08,coombes11}. In section \ref{inhom}, we study how periodic heterogeneities affect the stability and evolution of bumps in the presence of noise.

The nonlinearity $f$ is a firing rate function which converts synaptic inputs $u$ to a resulting fraction of active neurons, between zero and one by definition. In line with experimental observations, this is often taken to be a sigmoidal function \cite{wilson73,coombes05,bressloff12}
\begin{align}
f(u) &= \frac{1}{1 + \e^{- \gamma ( u - \theta )}},   \label{sig}
\end{align}
where $\gamma$ is the gain and $\theta$ is the threshold. We can perform much of our analysis for a general firing rate function $f$, such as the sigmoid (\ref{sig}). However, one particular idealization that eases mathematical analysis considers the infinite gain $\gamma \to \infty$ limit, so that (\ref{sig}) becomes a Heaviside step function \cite{amari77,coombes05,bressloff12}
\begin{align}
f(u) = H(u - \theta) = \left\{ \begin{array}{cl} 0 & : u< \theta, \\ 1 & : u \geq \theta. \end{array} \right. \label{H}
\end{align}
The Heaviside firing rate function (\ref{H}) allows us to explicitly calculate many quantities of interest in our study.

As mentioned, the deterministic neural field equation (\ref{ring}) has been studied extensively as a model of neural pattern formation \cite{benyishai97,camperi98,bressloff02,veltz10}. The main interest of this paper is to consider effects of fluctuations on stationary bump solutions of (\ref{ring}). In particular, we will consider the general case where noise can depend multiplicatively on the state variable $u$. Thus, we have the following Langevin equation that describes a noisy neural field
\begin{align}
\d \U (x,t) &= \left[ -\U (x,t) + \int_{- \pi}^{\pi} w(x,y) f( \U (y,t)) \d y \right] \d t + \ve^{1/2} g(\U (x,t)) \d W (x,t), \label{ringlang}
\end{align}
where $\U (x,t)$ tracks the sum of synaptic inputs at position $x \in ( - \pi, \pi)$ at time $t$. The term $\d W(x,t)$ is the increment of a spatially dependent Wiener process such that
\begin{align}
\langle \d W(x,t) \rangle = 0, \ \ \ \langle \d W (x,t) \d W(y,s) \rangle = C(x - y) \delta (t - s) \d t \d s, \label{wien}
\end{align}
so that $\ve$ determines the noise amplitude, which is weak ($\ve \ll 1$). Spatial correlations of the noise are described by the function $C(x-y)$, which is symmetric and depends on the distance between two spatial locations in the network. The function $g(\U)$, describing the multiplicative noise, can be specified arbitrarily in a great deal of our analysis. Models such as (\ref{ringlang}) have been recently introduced as stochastic versions of neural field equations \cite{laing01b,hutt08,faugeras09b,bressloff12b}. Note that we can also examine the effects of additive noise in the framework (\ref{ringlang}) by simply taking the function $g(\U) = 1$.

The paper is organized as follows. First, in section \ref{bump1} we study the effects of noise on bump solutions of the ring network (\ref{ring}) with spatially homogeneous synaptic weights $w(x,y) = \w (x - y)$. Since bumps have a zero eigenvalue associated with their linear stability, indicating marginal stability and translation symmetry, the introduction of noise in the Langevin equation (\ref{ringlang}) leads to their wandering as a purely diffusive process. We compute the effective diffusion coefficient of this process as well as a shift in the mean width of the bump due to the multiplicative noise. We examine the effects of spatially dependent inputs in section \ref{input}. Since, in this case, the network will no longer be (continuously) translationally symmetric, stable bumps are linearly stable to perturbations of their mean position. Introducing noise then leads to their position evolving as a mean-reverting stochastic process, rather than a purely diffusive one. However, on exponentially long time scales, the bump can escape from the position to which they are pinned so they move to the vicinity of another discrete attractor of the deterministic system. In section \ref{inhom}, we introduce spatially periodic heterogeneities into the weight function so that bumps still exist, but calculation of their stability reveals there is no longer a generic zero eigenvalue, once again reflecting the loss of translation symmetry. As in the case of external inputs, this leads to pinning of bumps to a finite number of discrete attractors in the stochastic system (\ref{ringlang}) so their position evolves as a mean-reverting process. Even though bumps can escape from these pinned positions, they ultimately wander with a smaller effective diffusion coefficient than in the spatially homogeneous network.

\section{Wandering bumps in spatially homogeneous network}
\label{bump1}
We begin by studying bumps in a spatially homogeneous ring model ($w(x,y) = \w (x-y)$), first in the absence of noise (\ref{ring}) and then with multiplicative noise (\ref{ringlang}). Previous studies of traveling fronts in reaction diffusion equations and neural fields have found multiplicative noise can alter the mean speed of the front and causes the front to wander diffusively \cite{armero98,panja04,sagues07,brackley07,bressloff12b}. Analyzing (\ref{ringlang}) reveals that multiplicative noise leads to dynamics whose mean is given by a bump with a position that wanders diffusively. Our analysis allows us to approximate the diffusion coefficient of the bump, estimating the error a network may make in a working memory task that relies on the position of the bump center \cite{white94,ploner98,camperi98,chow06}.

\subsection{Existence}
To begin, we derive stationary bump solutions. As opposed to the method of construction of Amari \cite{amari77}, we need not presume a Heaviside firing rate function (\ref{H}) to derive explicit bump solutions. We exploit the fact that the cosine weight function (\ref{cos}) is separable through a trigonometric identity to reduce the existence and stability problems to root-finding problems or linear algebraic systems \cite{hansel98,veltz10}. 

Upon assuming a stationary solution $u(x,t) = U(x)$, the scalar equation (\ref{ring}) with a spatially homogeneous weight function $\w(x-y)$ requires that it satisfy the integral equation
\begin{align}
U(x) &= \int_{- \pi}^{\pi} \w (x-y) f(U(y)) \d x. \label{bumpwf}
\end{align}
For the weight function (\ref{cos}), we can employ the trigonometric identity
\begin{align}
\cos(x-y) = \cos x \cos y + \sin x \sin y, \label{cosid}
\end{align}
so that (\ref{bumpwf}) becomes
\begin{align}
U(x) &= A \cos x + B \sin x,   \label{UbumpAB}
\end{align}
where
\begin{align}
A =  \int_{- \pi}^{\pi} \cos x f(U(x)) \d x, \ \ \ \ \ \ \ \ \  B  =  \int_{- \pi}^{\pi} \sin x f(U(x)) \d x. \label{bumpA}
\end{align}
We look specifically for even symmetric stationary bump solutions, as is often done in analyses of localized solutions in neural fields \cite{amari77,coombes05,veltz10,bressloff12}. Thus, $B=0$, so
\begin{align}
U(x) = A \cos x,  \label{bsolA}
\end{align}
and we can solve for $A$ by requiring self consistency of the solution $U = A \cos x$ such that (\ref{bumpA}) becomes
\begin{align}
A &=  \int_{- \pi}^{\pi} \cos x f(A \cos x ) \d x. \label{bAselfcon}
\end{align}
For a general sigmoidal firing rate function (\ref{sig}), one could determine $A$ using a numerical root finding method.

For a Heaviside firing rate function (\ref{H}), we can solve exactly for the amplitude $A$. Equation (\ref{bsolA}) shows $U(x)$ is unimodal and symmetric, so it will cross above and below $\theta$ at locations $x=-a$ and $x=a$ respectively. This provides us with the threshold conditions $U( \pm a ) = \theta$ for (\ref{bsolA}), which can be written equivalently as
\begin{align}
a = \cos^{-1} \frac{\theta}{A}.  \label{wid1beqn}
\end{align}
Thus, we know $U(x) > \theta$ for $x \in ( - \cos^{-1} ( \theta/ A), \cos^{-1} ( \theta/ A))$, so the self-consistency condition (\ref{bAselfcon}) becomes
\begin{align}
A &= 2  \int_0^{\cos^{-1} \theta/A} \cos x \d x = 2  \sin \left( \cos^{-1} \frac{\theta}{A} \right) = 2 \sqrt{1 - \frac{\theta^2}{A^2}}. \label{bAHeav}
\end{align}
Solving (\ref{bAHeav}) for the bump scaling factor
\begin{align}
A &= \sqrt{1+  \theta} \pm \sqrt{1 -  \theta}  \label{bdetscal}
\end{align}
reveals there are two bump solutions
\begin{align}
U(x)  &= \left(  \sqrt{1+  \theta} \pm \sqrt{1 -  \theta} \right) \cos x,  \label{bdetsol}
\end{align}
and we can show that the wide solution ($+$) is stable and the narrow solution ($-$) is unstable, forming a separatrix between the wide bump and the rest state $U(x) = 0$. Applying (\ref{wid1beqn}), half-widths $a$, can be easily computed
\begin{align}
a_{\pm} &= \cos^{-1} \left( \frac{\sqrt{1+  \theta} \mp \sqrt{1 -  \theta}}{2} \right). \label{b1widths}
\end{align}
As we have mentioned, the network with a cosine weight kernel (\ref{cos}) is translationally symmetric, so that we could construct a bump solution centered at any position $x \in [ - \pi , \pi ]$. This would simply lead to a system of two equations for $A$ and $B$ associated with (\ref{UbumpAB}), but the width of such a bump would be the same as that of (\ref{bsolA}). We can also show this by calculating the linear stability of bumps in the network (\ref{ring}), revealing marginal stability of a shift perturbation. This we do now for the case of a general firing rate function $f$.

\subsection{Stability}

Linear stability of bumps (\ref{bsolA}) can be computed by analyzing the evolution of small, smooth, separable perturbations such that $u(x,t) = U(x) + \psi (x) \e^{\lambda t}$ for $| \psi (x) | \ll 1$ . Plugging this expansion into the evolution equation (\ref{ring}), Taylor expanding, applying (\ref{bumpwf}), and studying first order equation yields \cite{bressloff02,coombes04,veltz10}
\begin{align}
( \lambda + 1 ) \psi (x) = \int_{- \pi}^{\pi} \w (x-y) f'(U(y)) \psi (y) \d y. \label{evalb1}
\end{align}
For the cosine weight function (\ref{cos}), we can apply the identity (\ref{cosid}) so that
\begin{align}
(\lambda + 1) \psi (x) = {\mc A} \cos x + {\mc B} \sin x, \label{beigcon}
\end{align}
where
\begin{align}
{\mc A} = \int_{- \pi}^{\pi} \cos x f'(U(x)) \psi (x) \d x, \hspace{1cm} {\mc B} = \int_{- \pi}^{\pi} \sin x f'(U(x)) \psi (x) \d x. \label{beig}
\end{align}
Thus, we reduce the infinite dimensional equation (\ref{evalb1}) to a $2 \times 2$ linear spectral problem (\ref{beigcon}). Such a technique was recently shown for a general class of weight functions in \cite{veltz10}. Plugging the form of $\psi (x)$ given by (\ref{beigcon}) into the system of equations (\ref{beig}), we have
\begin{align}
( \lambda + 1 ) \left( \begin{array}{c} {\mc A} \\ {\mc B} \end{array} \right) &= \left( \begin{array}{cc} {\mc I}(\cos^2 x) & {\mc I} ( \cos x \sin x ) \\ {\mc I} ( \cos x \sin x ) & {\mc I} ( \sin^2 x ) \end{array} \right)   \left( \begin{array}{c} {\mc A} \\ {\mc B} \end{array} \right),  \label{stab2b2}
\end{align}
where
\begin{align}
{\mc I} ( r(x) ) &= \int_{- \pi}^{\pi} r(x) f'(U(x)) \d x. \label{bintfp}
\end{align}
First of all, note that the essential spectrum is $\lambda = -1$ and thus does not contribute to any instabilities. Upon integrating (\ref{bAselfcon}) by parts, we see
\begin{align}
A = \int_{- \pi}^{\pi} \cos x f(A \cos x ) \d x = A \int_{- \pi}^{\pi} \sin^2 x f'(A \cos x ) \d x. \label{sin2fpid}
\end{align}
Therefore, as long as $A \neq 0$, the equality (\ref{sin2fpid}) tells us
\begin{align}
{\mc I} ( \sin^2 x) = \int_{- \pi}^{\pi} \sin^2 x f'(U(x)) \d x = 1.  \label{sin2id}
\end{align}
Using this identity (\ref{sin2id}) and the fact that (\ref{bintfp}) is linear, we can then compute
\begin{align}
{\mc I}( \cos^2 x ) = {\mc I}(1 - \sin^2 x) =  {\mc I}(1) - {\mc I}( \sin^2 x)  = {\mc I}(1) - 1. \label{Icosid}
\end{align}
Finally, we can use integration by parts to show
\begin{align}
{\mc I} ( \cos x \sin x ) = \int_{- \pi}^{\pi} \cos x \sin x f'(U(x)) \d x = - \int_{- \pi}^{\pi} \sin x f(U(x)) \d x = 0,  \label{csfpid}
\end{align}
since $U(x)$ is even. Using the identities (\ref{sin2id}), (\ref{Icosid}), and (\ref{csfpid}), it is straightforward to compute the eigenvalues that determine the stability of the bump (\ref{bsolA}). We do so by finding the roots of the associated characteristic equation
\begin{align*}
\lambda^2 + (2 - {\mc I}(1)) \lambda = 0,
\end{align*}
which reveals the zero eigenvalue $\lambda_o = 0$, associated with the constant ${\mc B}$, defined in (\ref{beig}), which means it reveals the linear stability of bumps in response to odd (shifting) perturbations. The fact that $\lambda_o$ is zero arises due to the underlying translation symmetry of (\ref{ring}) when $w(x,y)$ is the cosine weight function (\ref{cos}). In addition, the stability of the bump (\ref{bsolA}) is determined by the sign of the other eigenvalue
\begin{align}
\lambda_e = 2 \int_0^{\pi} f'(U(x)) \d x - 2, \label{nzevbump}
\end{align}
associated with ${\mc A}$, defined by (\ref{beig}), and thus even (expanding or contracting) perturbations of the bump.

In the limit of infinite gain $\gamma \to \infty$, $f$ becomes the Heaviside (\ref{H}), and
\begin{align}
f'(U(x)) = \frac{\d H(U(x))}{\d U} = \frac{\delta (x-a)}{|U'(a)|} + \frac{\delta (x +a)}{|U'(a)|},  \label{dH}
\end{align}
in the sense of distributions, so the formula for the generically nonzero eigenvalue will be
\begin{align}
\lambda_e = - 2 + \frac{2}{|U'(a)|}, \label{Hevbump}
\end{align}
for the bump (\ref{bdetsol}) of half-width $a$. Identifying threshold $\theta$ values at which (\ref{Hevbump}) crosses zero will give the location of a saddle-node bifurcation \cite{amari77,coombes04,folias04}. Equation (\ref{Hevbump}) allows us to compute eigenvalues exactly for the wide and narrow bumps since (\ref{b1widths}) gives the half-widths and the spatial derivative at the edges
\begin{align}
|U'(a_{\pm})| &= \left( \sqrt{1+ \theta} \pm \sqrt{1- \theta} \right) \sin \left( \cos^{-1} \left( \frac{\sqrt{1+ \theta} \mp \sqrt{1 - \theta}}{2} \right) \right) \nonumber \\
&=  \left( \sqrt{1+ \theta} \pm \sqrt{1- \theta} \right) \sqrt{\frac{1 \pm \sqrt{1- \theta^2}}{2}}. \label{Upabump}
\end{align}
Plugging the expression (\ref{Upabump}) into (\ref{Hevbump}) yields
\begin{align}
\lambda_e = \lambda_{\pm} = -2 + \frac{2 \sqrt{2}}{( \sqrt{1+ \theta} \pm \sqrt{1- \theta} ) \sqrt{1 \pm \sqrt{1- \theta^2}}}, \label{wdnarevbump}
\end{align}
the nonzero eigenvalue associated with the wide ($+$) and narrow ($-$) bump. To identify the threshold $\theta$ where the two pulses annihilate in a saddle-node bifurcation, we look for where $\lambda_{\pm} = 0$. Imposing this requirement on (\ref{wdnarevbump}) means
\begin{align}
\left( \sqrt{1+ \theta} \pm \sqrt{1- \theta} \right) \sqrt{1 \pm \sqrt{1- \theta^2}} = \sqrt{2}. \label{Hbsn1}
\end{align}
It can be shown that (\ref{Hbsn1}) is equivalent to finding zeros of the quartic $\theta^4 + 2 \theta^2 -3$, whose real solutions are $\theta = \pm 1$. Thus, as $\theta$ is increased from zero, the stable wide and unstable narrow bump branches will coalesce in a saddle-node bifurcation at $\theta = 1$.

\subsection{Noise-induced wandering of bumps}

We now consider the effect noise has on bumps by studying approximate solutions to the Langevin equation (\ref{ringlang}) with a spatially homogeneous weight function $w(x,y) = \w (x-y)$. The primary behavior in which we are interested is how the bump's position changes. Wandering of bumps was first observed numerically in modeling studies of working memory that employed rate \cite{camperi98} and spiking models \cite{compte00}. These authors rightly observed that such pure diffusion was due to the potential landscape of the deterministic dynamical system being a line attractor \cite{camperi98,brody03}. In the case of truly multiplicative noise, we show that the mean width of the bump changes changes as well. Mainly, we show that we can use a linear expansion to approximate the influence of spatially correlated multiplicative noise on the position of bumps in a neural field. This reveals that the bump undergoes pure diffusion whose associated coefficient we can derive from our asymptotic analysis.

The fact that multiplicative noise alters the mean width of the bump arises from the fact that this noise does not have zero mean, $\langle g(\U) \d W \rangle \neq 0$. We can calculate this average using Novikov's theorem \cite{novikov65,armero98,sagues07,bressloff12b}
\begin{align}
\ve^{1/2} \langle g( \U ) \d W \rangle  = \ve C(0) \langle g'( \U ) g( \U ) \rangle \d t. \label{novikov}
\end{align}
One method for deriving the result (\ref{novikov}) is to Fourier transform (\ref{ringlang}) and evaluate averages using the corresponding Fokker-Planck equation in Fourier space \cite{sagues07}. Rewriting equation (\ref{ringlang}) using (\ref{novikov}), we can formulate the fluctuating term so that it has zero mean
\begin{align}
\d \U (x,t) = \left[ -h( \U (x,t)) + \int_{- \pi}^{\pi} \w(x - y) f ( \U (y,t)) \d y \right] \d t + \ve^{1/2} \d Z ( \U , x, t),  \label{rlangzm}
\end{align}
where
\begin{align}
h(\U (x,t)) = \U (x,t)  - \ve C(0) g'( \U ( x,t) ) g( \U ( x,t) )  \label{hfblang}
\end{align}
and
\begin{align}
\d Z ( \U ,x,t) = g( \U ) d W(x,t) - \ve^{1/2} C(0) g'( \U ) g( \U ) \d t .   \label{dZnoise}
\end{align}
The stochastic process $Z$ has zero mean and variance
\begin{align}
\langle dZ ( \U ,x,t) \d Z ( \U , y,t) \rangle = \langle g( \U (x,t)) \d W(x,t) g( \U (y,t)) \d W (y,t) \rangle + {\mc O} ( \ve^{1/2} ).
\end{align}
Next, we assume that the multiplicative noise in (\ref{rlangzm}) generates two phenomena that occur on disparate time scales. Diffusion of the bump from its original position occurs on long timescales, and fluctuations in the bump profile occur on short time scales \cite{mikhailov83,armero98,bressloff12b}. Thus, we express the solution $\U$ of equation (\ref{rlangzm}) as the sum of a fixed bump profile $U_{\ve}$ displaced by $\Delta (t)$ from its mean position $x$, and higher order time-dependent fluctuations $\ve^{1/2} \Phi + \ve \Phi_1 + \ve^{3/2} \Phi_2 + \cdots$ in the profile of the bump
\begin{align}
\U (x,t) = U_{\ve} (x - \Delta (t)) + \ve^{1/2} \Phi ( x- \Delta (t)  , t) + \cdots ,  \label{ringbexp}
\end{align}
so $\Delta (t)$ is a stochastic variable indicating the displacement of the bump $U_{\ve}$ with slightly altered half-width $a_{\ve}$. To a linear approximation, the stochastic variable $\Delta (t)$ undergoes pure diffusion with associated coefficient $D( \ve ) = {\mc O} ( \ve )$, as we show. The expansion (\ref{ringbexp}) is not a standard small-noise expansion, since the modified bump $U_{\ve}$ implicitly depends upon $\ve$, where the subscript denotes parametrization. By substituting (\ref{ringbexp}) into equation (\ref{rlangzm}) and taking averages, we find the leading order deterministic equation for $U_{\ve}$ is
\begin{align}
h(U_{\ve} (x)) = \int_{- \pi}^{\pi} \w (x-y) f(U_{\ve} (y)) \d y.  \label{langmbump}
\end{align}
The mean pulse half-width $a_{\ve}$ and profile $U_{\ve} $ depend non-trivially on the noise strength $\ve$, since $h$, given by (\ref{hfblang}), is $\ve$-dependent. Therefore, the width $a_{\ve} \neq a$ for $\ve > 0$ and $a_{0} = a$, the width of the bump in the absence of multiplicative noise. Proceeding to next order, and requiring (\ref{langmbump}), we find $\Delta (t) = {\mc O}( \ve ^{1/2} )$ and
\begin{align}
\d \Phi (x,t) = {\mc L} \Phi (x,t) + \ve^{-1/2} U_{\ve} '( x) \d \Delta (t) + \d Z (U_{\ve}, x, t), \label{dPlangb}
\end{align}
where ${\mc L}$ is the non-self-adjoint linear operator
\begin{align}
{\mc L} p (x) = -h'(U_{\ve} (x)) p(x) + \int_{- \pi}^{\pi} \w (x -y) f'(U_{\ve}(y)) p(y) \d y,  \label{bmlinop}
\end{align}
for any function $p(x) \in L^2 [- \pi, \pi ] $.

Upon differentiating (\ref{langmbump}) and integrating the convolution by parts
\begin{align*}
h'(U_{\ve} (x)) U_{\ve}'(x) = \int_{- \pi}^{\pi} \w (x-y)f'(U_{\ve}(y))U_{\ve}'(y) \d y,
\end{align*}
so $U_{\ve}'(x)$ belongs to the nullspace of ${\mc L}$. Now, we can ensure a bounded solution to equation (\ref{dPlangb}) exists by requiring the inhomogeneous part is orthogonal to all elements of the nullspace of the adjoint operator ${\mc L}^*$. The adjoint is defined with respect to the $L^2$ inner product
\begin{align*}
\int_{- \pi}^{\pi} \left[ {\mc L} p(x) \right] q(x) \d x = \int_{- \pi}^{\pi} p(x) \left[ {\mc L}^* q(x) \right] \d x,
\end{align*}
where $p(x), q(x) \in L^2 [ - \pi , \pi ] $. Thus,
\begin{align}
{\mc L}^* q(x) = -h'(U_{\ve}(x)) q(x) + f'(U_{\ve} (x)) \int_{- \pi}^{\pi} \w (x-y) q(y) \d y.  \label{blangadjop}
\end{align}
There is a single function $\vp_{\ve} (x)$ spanning the one-dimensional nullspace of ${\mc L}^*$, which we can compute explicitly for a general firing rate function $f$. Thus, we impose solvability of (\ref{dPlangb}) by taking the inner product of both sides of the equation with respect to $\vp_{\ve}(x)$ yielding
\begin{align*}
\int_{- \pi}^{\pi} \vp_{\ve} (x) \left[ U_{\ve} '(x) \d \Delta (t) + \ve^{1/2} \d Z(U_{\ve}, x,t) \right] \d z = 0.
\end{align*}
Isolating $\d \Delta (t)$, we find $\Delta (t)$ satisfies the stochastic differential equation (SDE)
\begin{align}
\d \Delta (t) = - \ve^{1/2} \frac{\D \int_{- \pi}^{\pi} \vp_{\ve} ( x)  \d Z(U_{\ve} ,x,t) \d x}{\D \int_{- \pi}^{\pi} \vp_{\ve} (x) U_{\ve}'(x) \d x}. \label{sdebumdel}
\end{align}
With the SDE (\ref{sdebumdel}) in hand, we can compute the effective diffusivity of the bump to a linear approximation. First, note that the mean position of the bump averaged over realizations does not change in time
\begin{align*}
\langle \Delta (t) \rangle = - \ve^{1/2} \frac{\int_{- \pi}^{\pi} \vp_{\ve}(x) \langle Z ( U_{\ve}, x,t) \rangle \d x}{\int_{- \pi}^{\pi} \vp_{\ve} (x) U_{\ve}'(x) \d x} t = 0,
\end{align*}
where we set the bump's initial position to be $\Delta (0) = 0$ without loss of generality. Computing the variance of the stochastic variable $\Delta (t)$, we find it evolves according to pure diffusion since
\begin{align}
\langle \Delta(t)^2 \rangle &= \ve \frac{\int_{- \pi}^{\pi} \int_{- \pi}^{\pi} \vp_{\ve} (x)  \vp_{\ve} ( y) g ( U_{\ve} ( x)) g(U_{\ve}(y)) \langle W(x,t) W(y,t) \rangle \d y \d x}{\left[ \int_{- \pi}^{\pi} \vp_{\ve} (x) U_{\ve}'(x) \d x \right]^2}t  \nonumber \\ 
\langle \Delta (t)^2 \rangle &= D( \ve ) t,    \label{bvarlin}
\end{align}
and using the definition of $W(x,t)$ in (\ref{wien}) yields
\begin{align}
D( \ve ) &= \ve \frac{\int_{- \pi}^{\pi} \int_{- \pi}^{\pi} \vp_{\ve} ( x) \vp_{\ve} ( y) g( U_{\ve} (x)) g(U_{\ve}(y)) C(x-y) \d y \d x}{\left[ \int_{- \pi}^{\pi} \vp_{\ve} (x) U_{\ve}'(x) \d x \right]^2} .   \label{bdiffcoeff}
\end{align}
To calculate the diffusion coefficient $D( \ve)$ for specific cases, we need to compute the constituent functions $U_{\ve}'(x)$ and $\vp_{\ve} (x)$. We will continue to use the cosine weight kernel (\ref{cos}) in this analysis.

\subsection{Calculating the diffusion coefficient}

First, we study additive noise, where $g(\U) = 1$, performing calculations for a general firing rate function $f$. For additive noise, the modification to the fluctuating term in the Langevin equation (\ref{ringlang}) using (\ref{novikov}) and (\ref{hfblang}) is not necessary because $\langle g'( \U ) g( \U ) \rangle = 0$ so $h(\U ) = \U$. Also, we find that equation (\ref{langmbump}) becomes (\ref{bumpwf}), so that 
\begin{align}
U_{\ve}' ( x) = U'(x)  = -A \sin x,   \label{baddspat}
\end{align}
where $A$ is defined by (\ref{bAselfcon}). Along these lines, the nullspace $\vp_{\ve} (x)$ of the adjoint ${\mc L}^*$ will depend trivially on $\ve$ as well so $\vp_{\ve} (x) = \vp (x)$. To find $\vp (x)$ in the case of additive noise, we write the (\ref{blangadjop}) using $h(\U) = \U$ and the kernel $\w (x) = \cos x$ so
\begin{align}
\vp (x) = f'(U(x)) \int_{- \pi}^{\pi} \cos ( x - y ) \vp ( y) \d y.  \label{nullspadd}
\end{align}
Using separability of the cosine kernel (\ref{cosid}), we find $\vp$ must satisfy
\begin{align}
\vp (x) = C f'(U(x)) \cos x + S f'(U(x)) \sin x,  \label{vpadd}
\end{align}
where
\begin{align}
C = \int_{- \pi}^{\pi} \cos x \vp (x) \d x, \hspace{2cm} S = \int_{- \pi}^{\pi} \sin x \vp (x) \d x. \label{CSadd}
\end{align}
Plugging the expression (\ref{vpadd}) into the pair of equations (\ref{CSadd}) gives us the linear system
\begin{align}
C &= {\mc I}( \cos^2 x ) C + {\mc I}( \cos x \sin x ) S, \nonumber \\
S &= {\mc I}( \cos x \sin x ) C + {\mc I} ( \sin^2 x ) S, \label{csystadd}
\end{align}
where ${\mc I}(r(x))$ is given by (\ref{bintfp}). Upon applying the identities (\ref{sin2id}), (\ref{Icosid}), and (\ref{csfpid}), the system (\ref{csystadd}) becomes
\begin{align*}
2 C = {\mc I}(1) C, \hspace{3cm}  S = S.
\end{align*}
As the gain $\gamma$ and threshold $\theta$ of the sigmoid (\ref{sig}) are varied ${\mc I}(1) \neq 2$ almost everywhere in $(\gamma, \theta )$. Thus, the only non-trivial solution to (\ref{nullspadd}) consistent across parameter values requires $C=0$, so
\begin{align}
\vp (x) = f'(U(x)) \sin x, \label{vpadsol}
\end{align}
up to the scaling $S$. Thus, for a general sigmoid (\ref{sig}), we can use our formula for the spatial derivative (\ref{baddspat}) along with (\ref{vpadsol}) to compute the term in the denominator of the diffusion coefficient (\ref{bdiffcoeff}) given
\begin{align*}
\int_{- \pi}^{\pi} \vp (x) U'(x) \d x = - A \int_{- \pi}^{\pi} \sin^2 x f'(U(x)) \d x = - A,
\end{align*}
applying (\ref{sin2id}). Thus, the effective diffusion coefficient in the case of additive noise is given
\begin{align}
D( \ve ) &= \frac{\ve}{A^2} \int_{- \pi}^{\pi} \int_{- \pi}^{\pi} \sin x \sin y f'(U(x)) f'(U(y)) C(x-y) \d y \d x.  \label{dcadd}
\end{align}

\begin{figure}
\begin{center} \includegraphics[width=12.5cm]{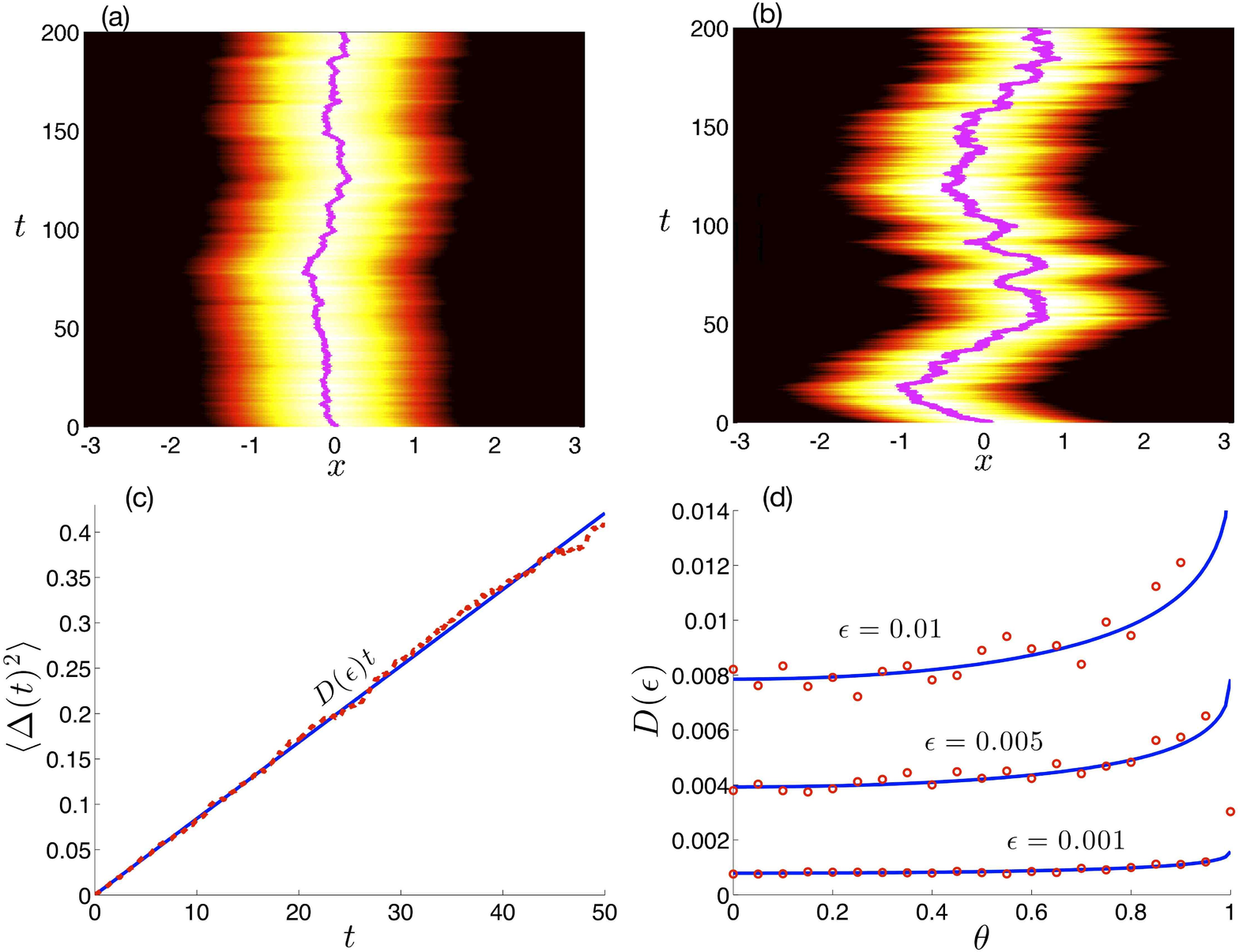}  \end{center}

\caption{Wandering of bumps due to additive noise with cosine correlation function (\ref{coscorr}), such that $g(\U) = 1$, in ring model (\ref{ringlang}) with Heaviside firing rate function (\ref{H}) and cosine weight kernel (\ref{cos}). (a) Single realization of neural activity $\U (x,t)$ driven by additive noise with amplitude $\ve = 0.001$, using stable stationary bump (\ref{bdetsol}) as initial condition. Superimposed line tracks center position (peak) of bump. Threshold $\theta = 0.5$. (b) Bump wanders more for higher amplitude noise $\ve = 0.01$. (c) Variance $\langle \Delta (t)^2 \rangle$ of bump center position computed across 1000 realizations (red dashed) scales linearly with time, as predicted by theory (blue solid). Diffusion coefficient $D( \ve )$ computed using (\ref{badddeH}). Parameters $\theta = 0.5$ and $\ve = 0.01$. (d) Dependence of diffusion coefficient of network threshold $\theta$ for $\ve = 0.001$ and $\ve = 0.01$ computed using asymptotic approximation (\ref{badddeH}) (blue line) and computed numerically (red circles) across 1000 realizations run for 50 time units. Numerical simulations of (\ref{ringlang}) are performed using Euler-Maruyama with a trapezoidal rule for the integral with the discretization $\Delta x = 0.01$ and $\Delta t = 0.01$.}
\label{baddwander}
\end{figure}

To determine the diffusion coefficient (\ref{dcadd}), we must specify correlation function $C(x-y)$. Two limits, spatially homogeneous and spatially uncorrelated noise, will help us understand how the spatial profile of the noise affects diffusion. In the limit of spatially homogeneous correlations ($C(x-y) \equiv C_0$), the neural field specified by (\ref{ringlang}) is driven by a spatially homogeneous Wiener process $\d W_0 (t)$. In this case, the bump will not diffuse at all since (\ref{dcadd}) simplifies to
\begin{align*}
D( \ve ) = \frac{\ve C_0}{A^2} \left[ \int_{- \pi}^{\pi} \sin x f'(U(x)) \d x \right]^2 = 0,
\end{align*}
since $f'(U(x))$ is even. Only the width of the bump will fluctuate, which is not tracked by our first order approximation.

In the limit of no spatial correlations ($C(x-y) \to \delta (x- y)$), every spatial point receives noise from an identically distributed independent Wiener process.\footnote{One important fact to note is that if we attempt to numerically simulate (\ref{ringlang}) with spatially uncorrelated noise on a spatial mesh of width $\Delta x$, a nonzero correlation length $\Delta x$ arises from the discretization \cite{armero98,bressloff12}.} In this case, we can simplify (\ref{dcadd}) to find
\begin{align*}
D( \ve ) = \frac{\ve}{A^2} \int_{- \pi}^{\pi} \sin^2 x \left[ f'(U(x)) \right]^2 \d x,
\end{align*}
which is nonzero for $\ve > 0$. 

Now, to compare our asymptotic analysis to numerical simulations, we will study the effect of a cosine spatial correlation function
\begin{align}
C(x-y) = \pi \cos ( x - y ).  \label{coscorr}
\end{align}
The $\pi$ scaling factor arises when we compute the correlation function from the spatial noise filter given by an unscaled cosine ${\mc F}(x) = \cos x$. To see this we take a spatially uncorrelated Wiener process $\d \Upsilon (x,t)$ and filter it with a cosine to give
\begin{align*}
\d W(x,t) = \int_{- \pi}^{\pi} {\mc F} (x - y) \d \Upsilon (y,t) \d y = \int_{- \pi}^{\pi} \cos ( x-y) \d \Upsilon (y,t) \d y, 
\end{align*} 
where $\langle \d \Upsilon (x,t) \rangle = 0$ and $\langle \d \Upsilon (x,t) \d \Upsilon ( y, s) \rangle = \delta (x-y) \delta (t-s) \d t \d s$. Then the cross correlation of $\d W(x,t)$ is given
\begin{align*}
\langle \d W(x,t) \d W(y,s) \rangle &= \int_{- \pi}^{\pi} \int_{- \pi}^{\pi} \cos (x - x') \cos (y - y') \langle \d \Upsilon (x',t) \d \Upsilon (y',s) \rangle \d x' \d y'  \\
&= \int_{- \pi}^{\pi} \cos (x - x') \cos ( y - x') \d x' \delta (t - s) \d t \d s \\
\langle \d W(x,t) \d W(y,s) \rangle &= C(x - y) \delta (t - s) \d t \d s = \pi \cos ( x- y) \delta (t - s) \d t \d s,
\end{align*}
as given by (\ref{coscorr}). Therefore, in the case of cosine spatial correlations (\ref{coscorr}), the diffusion coefficient in the presence of additive noise (\ref{dcadd}) becomes
\begin{align}
D( \ve ) &= \frac{\ve \pi }{A^2} \int_{- \pi}^{\pi} \int_{- \pi}^{\pi} \sin x \sin y f'(U(x)) f'(U(y))  \cos (x-y) \d y \d x \label{badddesig} \\
&= \frac{\ve \pi}{A^2} \left[ \left( \int_{- \pi}^{\pi} \sin^2 x f'(U(x)) \d x \right)^2 + \left( \int_{- \pi}^{\pi} \sin x \cos x f'(U(x)) \d x \right)^2 \right] = \frac{\ve \pi}{A^2},   \nonumber
\end{align}
where we have applied the identities (\ref{cosid}), (\ref{sin2id}), and (\ref{csfpid}). In the case of a Heaviside firing rate function (\ref{H}), we can use the explicit expression (\ref{bdetscal}) for the amplitude of the stable bump to write (\ref{badddesig}) simply in terms of the noise amplitude $\ve$ and network threshold $\theta$ as
\begin{align}
D( \ve ) &= \frac{\ve \pi}{2 + 2 \sqrt{1 - \theta^2}}. \label{badddeH}
\end{align}
Thus, we have an asymptotic approximation for the effective diffusion coefficient $D( \ve )$ of a stable bump (\ref{bdetsol}) in the ring network (\ref{ringlang}) driven by additive noise, $g( \U ) = 1$. We compare (\ref{badddeH}) to diffusion coefficients computed from numerical simulations in Fig. \ref{baddwander}. As predicted by our theory, averaging across numerical realization the Langevin equation (\ref{ringlang}) shows the variance of the bump's position scales linearly in time.

Now, we examine effects of multiplicative noise. The main difference between this case and that of purely additive noise is that the mean width and amplitude of the bump are altered, as suggested by (\ref{langmbump}). In the case of a cosine weight kernel (\ref{cos}) and a linear multiplicative function $g( \U ) = \U $, the modified equation (\ref{langmbump}) for the bump becomes
\begin{align}
U_{\ve} (x) = \frac{1}{( 1 - \ve C(0))} \int_{- \pi}^{\pi} \cos y f( U_{\ve} (y)) \d y \cos x = A_{\ve} \cos x.  \label{bmmodeq}
\end{align}
The final equality of (\ref{bmmodeq}) shows that the form of the bump solution is the same as in noise-free case (\ref{bsolA}); only the amplitude is changed. Equation (\ref{bmmodeq}) can then give us a nonlinear equation specifying the amplitude $A_{\ve}$ according to
\begin{align}
A_{\ve} = \frac{1}{(1 - \ve C(0)) } \int_{- \pi}^{\pi} \cos x f ( A_{\ve} \cos x ) \d x. \label{bmnmodA}
\end{align}
As in the noise-free case, this equation is much easier to solve than the nonlinear integral equation (\ref{bmmodeq}). This also provides the spatial derivative for the mean bump profile
\begin{align}
U_{\ve}'(x) = - A_{\ve} \sin x.  \label{bmulsd}
\end{align}
Notice, by integrating the right hand side of (\ref{bmnmodA}) by parts and canceling $A_{\ve}$, as long as $A_{\ve} \neq 0$, we have the formula
\begin{align}
1 - \ve C(0) = \int_{- \pi}^{\pi} \sin^2 x f'(U_{\ve} (x) \d x = {\mc I}_{\ve} ( \sin^2 x),  \label{Ipsin2id}
\end{align}
where
\begin{align}
{\mc I}_{\ve} ( r(x) ) = \int_{- \pi}^{\pi} r(x) f'(U_{\ve} (x)) \d x. \label{Imnosip}
\end{align}
Notice, (\ref{Ipsin2id}) is analogous to the identity (\ref{sin2id}). As in that case, we can derive
\begin{align}
{\mc I}_{\ve} ( \cos^2 x ) = {\mc I}(1) + \ve C(0) - 1,  \ \ \ \ \  {\mc I}_{\ve} ( \cos x \sin x ) = 0.  \label{Ipnoscosid}
\end{align}

With the modified bump solution (\ref{bmmodeq}) in hand, the $\ve$-dependent nullspace $\vp_{\ve} (x)$ of the adjoint operator (\ref{blangadjop}) is given by
\begin{align}
(1 - \ve C(0)) \vp_{\ve} (x) = f'(U_{\ve} (x)) \int_{- \pi}^{\pi} \cos (x - y ) \vp_{\ve} (y) \d y.  \label{bmnosadje}
\end{align}
Following our analysis in the case of additive noise, we use the identity (\ref{cosid}) of the cosine kernel (\ref{cos}) to derive a linear system specifying $\vp_{\ve} (x)$ and apply the identities  (\ref{Ipsin2id}) and (\ref{Ipnoscosid}) to yield the solution to (\ref{bmnosadje}) given
\begin{align}
\vp_{\ve} (x) = f'(U_{\ve} (x)) \sin x.  \label{ansmnosb}
\end{align}
We then can use the spatial derivative (\ref{bmulsd}) and the nullspace formula (\ref{ansmnosb}) to compute the term in the denominator of the diffusion coefficient (\ref{bdiffcoeff}) given by
\begin{align*}
\int_{- \pi}^{\pi} \vp_{\ve} (x) U_{\ve}'(x) \d x = - A_{\ve} \int_{- \pi}^{\pi} \sin^2 x f'(U_{\ve} (x)) \d x = - A_{\ve} (1- \ve C(0)),
\end{align*}
where we have applied (\ref{Ipsin2id}). Thus, the effective diffusion coefficient in the case of multiplicative noise with $g(\U) = \U$ becomes
\begin{align}
D( \ve ) = \frac{\ve}{(1 - \ve C(0))^2} \int_{- \pi}^{\pi} \int_{- \pi}^{\pi} \cos x \sin x f'(U_{\ve}(x)) \cos y \sin y f'(U_{\ve}(y)) C(x - y) \d y \d x.  \label{bmdiffcoeff}
\end{align}

\begin{figure}
\begin{center} \includegraphics[width=12.5cm]{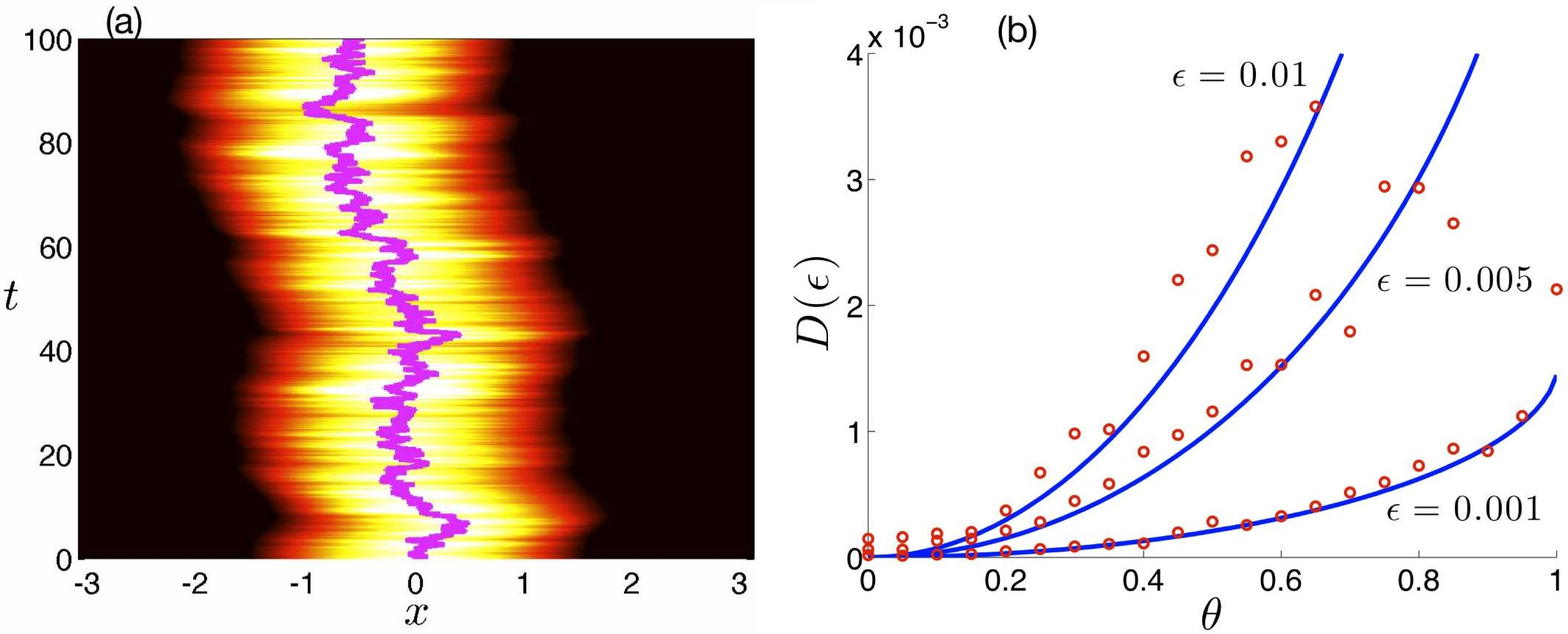} \end{center}
\caption{Wandering of bumps induced by multiplicative noise, such that $g(\U) = \U$, in ring model (\ref{ringlang}) with Heaviside firing rate function (\ref{H}) and cosine weight kernel (\ref{cos}). (a) Single realization of $\U (x,t)$ driven by multiplicative noise with amplitude $\ve = 0.01$, using stable stationary bump (\ref{bdetsol}) as initial condition. Superimposed line tracks center (peak) of bump. Threshold $\theta = 0.5$. (b) Dependence of diffusion coefficient of network threshold $\theta$ computed using asymptotic approximation (\ref{badddeH}) (blue solid) and computed numerically (red circles) across $1000$ realizations for $100$ time units. Numerical scheme is same as in Fig. \ref{baddwander}.}
\label{bmultwander}
\end{figure}

As we found in the case of additive noise, when noise is spatially homogeneous ($C(x) = C_0$), the diffusion coefficient defined by (\ref{bmdiffcoeff}) is $D( \ve ) = 0$. However, for $\ve >0$, the mean profile of the bump is modified, due to the $C(0) = C_0$ dependence of the modified bump amplitude equation (\ref{bmnmodA}). We cannot study the case of spatially uncorrelated noise (($C(x) \to \delta (x)$)) analytically, due to the delta distribution singularity ($C(0) \to \infty$) appearing in our explicit formulae for the amplitude $A_{\ve}$ (\ref{bmnmodA}) and diffusion coefficient $D( \ve )$ (\ref{bmdiffcoeff}). Thus, we proceed to analyze the case of spatially structured correlations.

In the case a cosine profile of spatial correlations (\ref{coscorr}), we can use the identity (\ref{cosid}) to find that (\ref{bmdiffcoeff}) becomes
\begin{align}
D ( \ve ) &= \frac{\ve \pi}{(1 - \ve \pi )^2} \left( \int_{- \pi}^{\pi} \sin^2 x \cos x f'(U_{\ve} (x)) \d x \right)^2, \label{bdifmcos}
\end{align}
where a second integral term vanishes, due to its integrand being odd. In the case of a Heaviside firing rate function (\ref{H}), we can compute the diffusion coefficient (\ref{bmdiffcoeff}) explicitly. To start with, we find the nonlinear equation (\ref{bmnmodA}) for the amplitude of the modified bump solution
\begin{align*}
A_{\ve} = \frac{\sqrt{1 + \theta (1 - \pi \ve )} + \sqrt{1 - \theta ( 1- \pi \ve )}}{1 - \pi \ve}. 
\end{align*}
This implies that the modified bump half-width $a = \cos^{-1} ( \theta/ A_{\ve})$ is
\begin{align}
a_{\ve} &= \cos^{-1} \left( \frac{\sqrt{1 + (1 - \pi \ve ) \theta} - \sqrt{1 - (1 - \pi \ve ) \theta}}{2} \right).   \label{bwmultH}
\end{align}
Now, we can compute the diffusion coefficient (\ref{bdifmcos}) explicitly
\begin{align}
D( \ve ) & = \frac{\ve \pi}{(1 - \ve \pi)^2} \left( \frac{\sin^2 a_{\ve} \cos a_{\ve} }{A_{\ve} \sin a_{\ve}} \right)^2 = \ve \pi \cos^2 a_{\ve} = \frac{\ve \pi (1 - \pi \ve )^2 \theta^2}{2 + 2 \sqrt{1 - (1 - \pi \ve )^2 \theta^2 }}.  \label{bmultdeH}
\end{align}
We compare our asymptotic estimation of the effective diffusion coefficient (\ref{bmultdeH}) to the results of numerical simulations in Fig. \ref{bmultwander}. Notice the different scaling of the diffusion coefficient as compared to that in the case of additive noise (\ref{badddeH}), especially in the vicinity of $\theta \approx 0$. This is due to the fact that lower network thresholds lead to there being weaker noise near the edges of the bump in the case of multiplicative noise. 

\subsection{Extinction of bumps near a saddle-node}

In general, there are few analyses that approximate the waiting times of large deviations in spatially extended systems with noise \cite{faris82,sowers95}. Recently, the approach of calculating minimum energy of the potential landscape of such systems has been used as a means of approximating the path of least action, along which a rare event is most likely to occur \cite{ren04}. Here, we show an example of a large deviation in the stochastic neural field (\ref{ringlang}) where the dynamics escapes from the basin of attraction of the stationary bump solution (\ref{bsolA}).

\begin{figure}
\begin{center} \includegraphics[width=12.5cm]{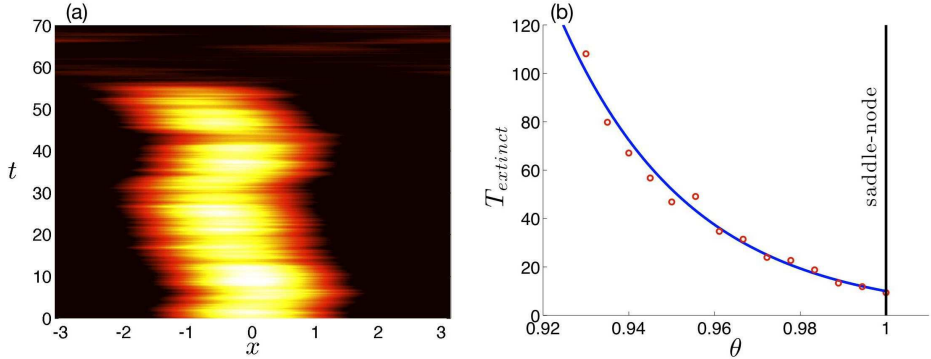}  \end{center}
\caption{Extinction of bumps in the network (\ref{ringlang}) with Heaviside firing rate function (\ref{H}) and additive noise with cosine spatial correlations (\ref{coscorr}). (a) Numerical simulation of (\ref{ringlang}) in presence of additive noise ($g(\U) = 1$), with threshold $\theta = 0.95$ and noise amplitude $\ve = 0.01$, where noise causes bump extinction at $t \approx 65$. (b) Numerical approximations (red circles) to the mean bump extinction time $T_{extinct}$ across 1000 realizations, given by when the bump's peak crosses below threshold $\theta$. This is fit to the exponential function $b \exp (\gamma |\theta - \theta_{SN}|)$ of the distance to the saddle-node at $\theta = \theta_{SN}$ using least squares (blue line). Specifically, $b \approx 10$ and $\gamma \approx 33$. Noise amplitude is $\ve = 0.01$. Numerical scheme is the same as in Fig. \ref{baddwander}.}
\label{extinctfig}
\end{figure}

We find that noise can cause trajectories of ${\mc U}(x,t)$ to cross through a separatrix of the deterministic system (\ref{ring}). This unstable manifold separates stable bump solutions from the homogeneous ``off" state. When multiplicative noise has no additive part ($g(0) = 0$), we expect the ${\mc U} \equiv 0$ state to be absorbing. In Fig. \ref{extinctfig}(a), we show the results of simulations where $g(\U)=1$, so that noise is purely additive. Here we take a Heaviside firing rate function (\ref{H}) and the threshold $\theta = 0.95$, so the system is operating near the saddle-node bifurcation of the deterministic system at $\theta_{SN}=1$ (see equation (\ref{Hbsn1})), and additive noise causes the bump to temporarily wander and then extinguish. Relating this to oculomotor delayed-response tasks, such an event would cause major error in the recall of a cue location. In Fig. \ref{extinctfig}(b), we show the mean time to extinction $T_{extinct}$ depends exponentially on the distance of the system to the saddle-node bifurcation, as described by the function $b \exp (\gamma |\theta - \theta_{SN}|)$. While we do not have a derivation of this formula per se, it stands to reason that the dynamics escapes some potential well whose height can be characterized by the distance $| \theta - \theta_{SN}|$. Thus, a Kramer's escape rate calculation could give the desired result \cite{gardiner04}. We will leave such analysis to future studies.

\section{Locking and escape of input-driven bumps}
\label{input}

Several studies of the ring model (\ref{ring}) have considered it to be an idealized model for the visual processing of oriented inputs \cite{hansel98,benyishai97,bressloff02,veltz10}. However, none of these have examined how well networks represent stimuli when they receive some source of noise. Here, we will study how well a network locks to a stationary stimulus, in the presence of purely additive noise
\begin{align}
\d \U (x,t) &= \left[ -\U (x,t) + \int_{- \pi}^{\pi} w(x-y) f( \U (y,t)) \d y + I(x) \right] \d t + \ve^{1/2} \d W (x,t). \label{inplang}
\end{align}
In the context of networks the encode working memories, an external input could be interpreted as feedback projections from another participating layer of neurons that may mirror the storage of (\ref{inplang}). Note, we could carry out an analogous analysis in the presence of multiplicative noise, but the formulation (\ref{inplang}) makes effects of input and noise more transparent. The term $I(x)$ represents a persistent external input to a network. For example, including
\begin{align}
I(x) = I_0 \cos nx  \label{cosn}
\end{align}
as our input turns the energy landscape of the deterministic ring network (\ref{ring}) from a line attractor (with continuous translation symmetry) to a chain of multiple attractors, such that the network now has dihedral $D_n$ rather than circular $O(2)$ symmetry. Note, any break in continuous translation symmetry, however weak, will considerably alter the governing dynamics of the deterministic system (\ref{ring}). In particular inputs can pin bumps in place so they do not wander freely.

\subsection{Existence of input-driven bumps}
Considering the deterministic version of the input-driven network (\ref{inplang}) with the $n$-modal input (\ref{cosn}), we first examine the stimulus driven bump solutions to the system
\begin{align}
\frac{\pd u(x,t)}{\pd t}  = - u(x,t) + \int_{- \pi}^{\pi} w(x - y) f(u(y,t)) \d y + I_0 \cos (nx). \label{ringcosn}
\end{align}
Looking for symmetric stationary solutions $u(x,t) = U(x)$, in the case of cosine weights (\ref{cos}), the equation (\ref{ringcosn}) becomes
\begin{align}
U(x) = \int_{- \pi}^{\pi} \cos y f(U(y)) \d y \cos x + I_0 \cos (nx).  \label{binpeqn}
\end{align}
Note, there is another class of solutions centered at $x = \pi/n$, rather than $x=0$. Since these solutions are always unstable for the ranges of parameters we are examining, we will ignore them for the time being. We will study such solutions in more detail in section \ref{inhom}, when we study a related system that considers spatial heterogeneity in synapses, rather than input. Thus, the solutions we study here will be of the form
\begin{align}
U(x) = A_1 \cos x + I_0 \cos (nx),  \label{bumpinp}
\end{align}
and we can write down an implicit equation for $A_1$ by requiring self consistency of the solution (\ref{bumpinp}), so the amplitude is given
\begin{align}
A_1 = \int_{- \pi}^{\pi} \cos y f (A_1 \cos y + I_0 \cos (ny) ) \d y, \label{binparbn}
\end{align}
and in the special case, $n=1$, we have
\begin{align}
A_1 = \int_{- \pi}^{\pi} \cos y f ( (A_1 + I_0) \cos y ) \d y.  \label{binpn1}
\end{align}
To demonstrate this analysis, we consider the case of a Heaviside firing rate function (\ref{H}). It is straightforward to evaluate the integral (\ref{binparbn}) using the fact that $U(x)$ is unimodal and thus only superthreshold in the region $x \in (-a,a)$, due to symmetry, so
\begin{align*}
A_1 = \int_{-a}^a \cos y \d y = 2 \sin a,
\end{align*}
then prescribing the threshold equation for self consistency yields
\begin{align}
2 \sin a \cos a + I_0 \cos na = \sin 2a + I_0 \cos na = \theta,  \label{bwidnarb}
\end{align}
which, in general, we can solve using a numerical root finding algorithm for the bump half-width $a$. Of course, as $n$ is increased, there are higher frequency modulations of the input (\ref{cosn}) that affect the profile of the bump. This can create problems in the requirement that the superthreshold region be a connected domain, that is that
\begin{align*}
U(x)> \theta \ : \  |x| < a \ \ \ \ \ {\rm and}  \ \ \ \ \ U(x) < \theta \ : \ |x|>a.
\end{align*}
Essentially, we need to guard against multibump solutions arising \cite{laing02,laing03}, as this complicates our analysis. In light of this, we restrict our study to small values of $n$ and $I_0$.

\begin{figure}
\begin{center} \includegraphics[width=12.5cm]{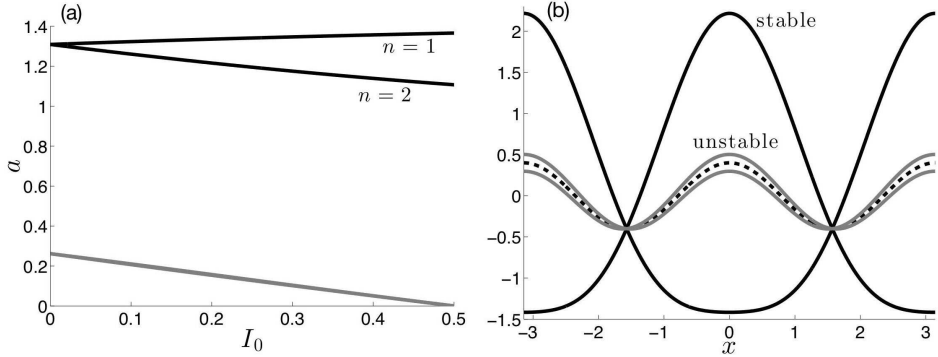} \end{center}
\caption{Input-locked bumps in the deterministic neural field (\ref{ringcosn}). (a) Bump half-width $a$ for a unimodal ($n=1$) and bimodal ($n=2$) stimulus of form (\ref{cosn}) calculated using (\ref{uniwidth}) and (\ref{biwidth}), demonstrating the dependence of the stable (black) and unstable (grey) branches on the input strength $I_0$. (b) Stable (black) and unstable (grey) bump solutions in the case of a bimodal stimulus (dashed) of strength $I_0 = 0.4$. Threshold is fixed at $\theta = 0.5$.}
\label{binpwidfig}
\end{figure}

To start, we note that, in the special case $n=1$, our equation for the bump half-width (\ref{bwidnarb}) becomes
\begin{align}
(2 \sin a + I_0 ) \cos a = \theta.  \label{uniwidth}
\end{align}
Because of the unimodality of the input $I(x) = I_0 \cos x$, we do not need to impose any additional super or subthreshold conditions, since the spatial frequency of the input ($n=1$) is commensurate with the spatial frequency of the bump. We demonstrate the dependence of the bump half-width $a$ upon the input strength $I_0$ in Fig. \ref{binpwidfig}(a). More interesting, however, is the implicit equation for the case $n=2$, given
\begin{align}
\sin 2a + I_0 \cos 2a = \theta.  \label{inpn2wid}
\end{align}
The equation (\ref{inpn2wid}) is explicitly solvable for the half-width $a$ in terms of parameters $\theta$ and $I_0$. To show this, we first multiply by $(I_0 + \theta)$ and apply half-angle formulae to yield
\begin{align*}
(I_0^2 + I_0 \theta ) ( \cos^2 a - \sin^2 a ) + 2 ( I_0 + \theta ) \cos a \sin a = \theta^2 + I_0 \theta.
\end{align*}
Upon rearranging terms, we can formulate the system as the equivalence of two squares
\begin{align*}
(1 - \theta^2 + I_0^2) \cos^2 a = \cos^2 a - 2 (I_0 + \theta ) \cos a \sin a + (I_0 + \theta)^2 \sin^2 a
\end{align*}
whose square roots can then be taken to yield
\begin{align*}
\pm \sqrt{1 - \theta^2 + I_0^2} \cos a = \cos a - (I_0 + \theta) \sin a.
\end{align*}
Then, upon rearranging terms, we can write the problem in terms of the level set crossings of the tangent function
\begin{align*}
\tan a = \frac{1 \pm \sqrt{1 - \theta^2 +I_0^2}}{I_0 + \theta},
\end{align*}
so that we can explicitly express the half-width $a$ of the bump in terms of parameters
\begin{align}
a = \tan^{-1} \left[ \frac{1 \pm \sqrt{1 - \theta^2 + I_0^2}}{I_0 + \theta} \right],  \label{biwidth}
\end{align}
where we restrict the range of $\tan^{-1}$ to yield $a \in [0, \pi]$. We demonstrate the dependence of the half-width $a$ on the input strength $I_0$ in Fig. \ref{binpwidfig}(a). In addition, we show how the profile is altered by a bimodal input in Fig. \ref{binpwidfig}(b). Now we turn to analyzing how inputs alter the stability of stationary bumps in the network.

\subsection{Stability of bumps locked to inputs}

As has been shown previous studies, stationary inputs can produce bumps that are linearly stable to translating perturbations, even though the input-free system is marginally stable to such perturbations \cite{benyishai97,bressloff02,folias04,veltz10}. Here, we demonstrate these results, as they provide intuition as to the alteration of the stochastic dynamics of the system (\ref{inplang}) from the input-free system (\ref{ringlang}). To do so, we perturb about $U(x)$ with small, smooth, separable functions using $u(x,t) = U(x) + \psi (x) \e^{\lambda t}$, where $|\psi (x)| \ll 1$, and perform a regular perturbation expansion, studying the first order equation
\begin{align}
( \lambda + 1) \psi (x) = \int_{- \pi}^{\pi} w(x-y) f'(U(y)) \psi (y) \d y, \label{oestabeq}
\end{align}
same as the input-free case (\ref{evalb1}) up to the different form of $U(x)$ given by (\ref{bumpinp}). In the case of a cosine weight function (\ref{cos}), we can separate the kernel $w$ in (\ref{oestabeq}), indicating that the function $\psi (x)$ will be of form (\ref{beigcon}). Therefore, as in the input free system, we can reduce the problem to a $2 \times 2$ linear system given by (\ref{stab2b2}). The main difference here is that the bump solution is given by (\ref{bumpinp}), as modified by the input. Therefore, upon calculating the eigenvalues associated with odd perturbations to the bump, we have
\begin{align}
\lambda_o = -1 + \int_{- \pi}^{\pi} \sin^2 y f'(A_1 \cos y + I_0 \cos (ny)) \d y.  \label{evaloddinp}
\end{align}
Upon integrating the right hand side of equation (\ref{binparbn}) by parts, we find that
\begin{align}
 \int_{- \pi}^{\pi} \sin^2 y f'(A_1 \cos y + I_0 \cos (ny)) \d y = 1 - \frac{n I_0}{A_1} \int_{- \pi}^{\pi} \sin y \sin (ny) f'(U(y)) \d y.  \label{sin2inp}
\end{align}
Therefore, using our equation for the eigenvalue $\lambda_o$ (\ref{evaloddinp}), the stability of the bump (\ref{bumpinp}) to odd perturbations will be determined by the sign of
\begin{align}
\lambda_o = - \frac{n I_0}{A_1} {\mc I}( \sin x \sin(nx)),  \label{lamboIeq}
\end{align}
where ${\mc I}(r(x))$ is given by (\ref{bintfp}). We are particularly interested in how the input (\ref{cosn}) alters the stability of of the bump to odd perturbations because the eigenvalue $\lambda_o \to 0$ in the limit $I_0 \to 0$. Therefore, infinitesimal changes in $I_0$ can alter the linear stability of the bump with respect to these perturbations. Using the formula (\ref{sin2inp}), we can also reduce the formula for the other eigenvalue
\begin{align*}
\lambda_e = -1 + \int_{- \pi}^{\pi} \cos^2 y f'(A_1 \cos y + I_0 \cos (ny)) \d y,
\end{align*}
associated with the even perturbations of the bump. This becomes
\begin{align}
\lambda_e = -2 + {\mc I}(1) + \frac{n I_0}{A_1} {\mc I}( \sin x \sin (nx)),  \label{lambeIeq}
\end{align}
whose sign will determine even perturbation stability.

To employ the linear stability theory we have developed, we study the case of a Heaviside firing rate function (\ref{H}). In this case, we know $A_1 = 2 \sin a$ and we can compute the integrals so that the eigenvalue formulae (\ref{lamboIeq}) and (\ref{lambeIeq}) reduce to
\begin{align}
\lambda_o = - \frac{n I_0 \sin (na)}{2 \sin^2 a + I_0 n \sin (na)}   \label{inpoddev}
\end{align}
and
\begin{align*}
\lambda_e = \frac{2 \cos (2a) -  n I_0 \sin (na)}{2 \sin^2 a + I_0 n \sin (na)}.
\end{align*}
Studying specific cases will help us understand how the input changes the stability of the bump (\ref{bumpinp}). In particular, if we start with the $n=1$ case, we have
\begin{align*}
\lambda_o = - \frac{I_0}{2 \sin a + I_0} < 0,
\end{align*}
since $a \in [0,\pi]$ by definition. Thus, an arbitrarily weak input will pin the bump (\ref{bumpinp}) to the position $x=0$ so that it is linearly stable to odd perturbations. The eigenvalue associated with even perturbations will be
\begin{align*}
\lambda_e = \frac{2 \cos (2a) - I_0 \sin a}{2 \sin^2 a + I_0 \sin a},
\end{align*}
whose sign is, in general, preserved from the input-free case ($I_0 = 0$). Moving to the $n=2$ case, the odd eigenvalue will be
\begin{align*}
\lambda_o = - \frac{I_0 \sin (2a)}{\sin^2 a + I_0 \sin ( 2a)},
\end{align*}
so that $\lambda_- < 0$ for sure when $a \in [0 , \pi / 2]$. The eigenvalue associated with even perturbations will be 
\begin{align*}
\lambda_e = \frac{\cos (2a) - I_0 \sin (2a)}{\sin^2 a + I_0 \sin (2a)}.
\end{align*}
In our analysis of the stochastic network (\ref{inplang}) with input (\ref{cosn}), we find the linear stabilization of odd perturbations to the bump allows it to remain pinned to a position, determined by the bump's center in (\ref{ringcosn}). This in contrast to the input-free system ($I_0=0$), in which the bump diffuses freely in the presence of noise.

\subsection{Mean-reverting stochastic process for bump location}

Now, we consider the effect of additive noise on the position of the bump in the stimulus-driven network (\ref{inplang}) with input (\ref{cosn}). Since the translation symmetry of the network has been broken, we find the stochastic variable describing bump location evolves as a mean-reverting (Ornstein-Uhlenbeck) process on moderate time scales. On very long timescales, large deviations occur, where the bump can escape from the vicinity of the stimulus peak at which it originally resided. 

To begin, we carry out a similar analysis to that of bumps evolving in the input free network (\ref{ringlang}). We express solutions to (\ref{inplang}) as a combination of the bump profile $U$ (same as the deterministic system) displaced by $\Delta (t)$ from its mean position, and an expansion of higher order time-dependent fluctuations $\ve^{1/2} \Phi + \ve \Phi_1 + \ve^{3/2} \Phi_2 + \cdots$ in the shape of the bump's profile so
\begin{align}
\U (x,t) = U(x - \Delta (t)) + \ve^{1/2} \Phi (x - \Delta (t), t) + \cdots.   \label{binpexp}
\end{align}
Substituting the expansion (\ref{binpexp}) into (\ref{inplang}) and taking averages, we find the leading order deterministic equation (\ref{binpeqn}), giving us the input-driven bump solution (\ref{bumpinp}). 

\begin{figure}
\begin{center} \includegraphics[width=12.5cm]{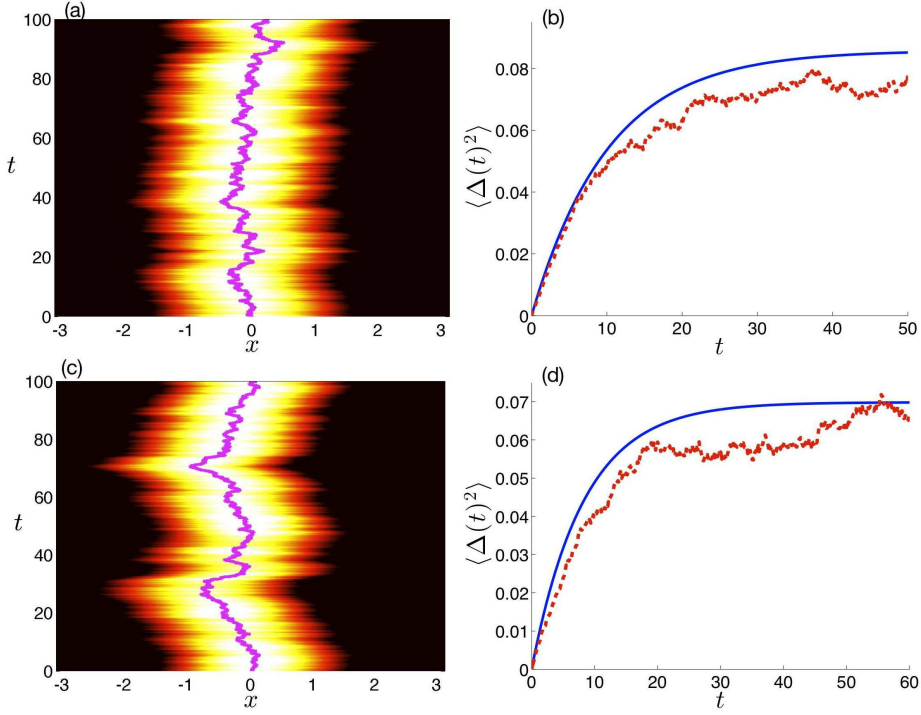} \end{center}
\caption{Bumps pinned by stationary inputs (\ref{cosn}) in the stochastic neural field (\ref{inplang}) with cosine correlated noise (\ref{coscorr}). (a) Numerical simulation for unimodal inputs ($n=1$). Bump stays in the vicinity of the stable fixed point at $x=0$. (b) Variance of the bump's position computed across 1000 realizations (red dashed) saturates, rather than growing linearly. Theoretical curve (blue solid), given by Ornstein-Uhlenbeck calculation (\ref{ouinput}) with (\ref{coskap}), compares nicely.  (c) Numerical simulation for bimodal inputs ($n=2$). Bump is initiated and stays in the vicinity of the fixed point at $x=0$, although there is another equilibrium of the deterministic system (\ref{ringcosn}) at $x= \pi$. (d) Variance of the bump's position for $n=2$. Other parameters are $\theta = 0.5$ and $\ve = 0.01$.}
\label{inppinfig}
\end{figure}

Proceeding to next order, we find $\Delta (t) = \Oo ( \ve^{1/2} )$ and
\begin{align}
\d \Phi (x,t) = \Lo \Phi (x,t) \d t + \ve^{-1/2} U'(x) \d \Delta (t) + \d W(x,t) + \ve^{-1/2} I' (x) \Delta (t) \d t ,  \label{dphinp}
\end{align}
where $\Lo$ is the non-self-adjoint operator
\begin{align}
\Lo p(x)  = -p(x) + \int_{- \pi}^{\pi} w(x-y) f'(U(y)) p(y) \d y,  \label{linopad}
\end{align}
for any function $p(x) \in L^2 [ - \pi, \pi ]$. Notice the last term on the right hand side of (\ref{dphinp}) arises due to the input. Since $U$ and $\Phi$ are functions of $x - \Delta (t)$, we have made the approximation $I(x) = I(x- \Delta (t) + \Delta (t)) \approx I( x - \Delta (t)) + I'(x - \Delta (t)) \Delta (t)$. Now, we can ensure that a bounded solution exists by requiring the inhomogeneous part of (\ref{dphinp}) is orthogonal to the nullspace $\vp (x)$ of the adjoint operator $\Lo^*$ defined by
\begin{align}
\Lo^* q (x) = -q(x) + f'(U(x)) \int_{ - \pi}^{\pi} w( x - y ) q(y) \d y. \label{adjopad}
\end{align}
Upon taking the $L^2$ inner product of both sides of (\ref{dphinp}) with $\vp (x)$ then provides a sufficient solvability condition
\begin{align}
\int_{- \pi}^{\pi} \vp (x) \left[ U'(x) \d \Delta (t) + I' (x) \Delta (t) \d t + \ve^{1/2} \d W(x,t) \right] \d x = 0.   \label{binpsolv}
\end{align}
We can rewrite (\ref{binpsolv}) to find that $\Delta (t)$ satisfies the SDE
\begin{align}
\d \Delta (t) + \kappa \Delta (t) \d t = \d {\mc W} (t),  \label{binpnsde}
\end{align}
where
\begin{align}
\kappa = \frac{\int_{- \pi}^{\pi} \vp (x) I' (x) \d x}{\int_{- \pi}^{\pi} \vp (x) U' (x) \d x}  \label{kapinp}
\end{align}
and
\begin{align}
{\mc W}(t) = - \ve^{1/2} \frac{\int_{- \pi}^{\pi} \vp (x) W(x,t) \d x}{\int_{- \pi}^{\pi} \vp (x) U'(x) \d x}. \label{whiteinp}
\end{align}
Note that the white noise term (\ref{whiteinp}) has the same diffusion coefficient as we computed in the input-free case for additive noise,
\begin{align*}
\langle \d {\mc W} (t) \rangle = 0, \ \ \ \ \ \ \ \  \ \langle \d {\mc W} (t) \d {\mc W} (t) \rangle = D ( \ve ) \d t
\end{align*}
where $D ( \ve )$ is given by (\ref{bdiffcoeff}) with $g( \U ) = 1$. Under the assumption that we begin the bump at a stable fixed point, we can calculate the mean and variance of the Ornstein-Uhlenbeck process using standard techniques \cite{gardiner04}
\begin{align}
\langle \Delta (t) \rangle = 0, \ \ \ \ \ \ \ \ \ \ \ \ \ \ \ \langle \Delta (t)^2 \rangle - \langle \Delta (t) \rangle^2 = \frac{D( \ve )}{2 \kappa} \left[ 1 - \e^{-2 \kappa t} \right].  \label{ouinput}
\end{align}
Thus, as opposed to the case of the freely diffusing bump, whose position's variance scales linearly with time as (\ref{bvarlin}), the stimulus-pinned bump's variance saturates at $D ( \ve ) / 2 \kappa$ in the large $t$ limit, according to (\ref{ouinput}). Variance saturation of bump attractors in networks with inputs has been demonstrated previously in simulations of spiking networks \cite{wu08}. Here, we have analytically demonstrated the mechanism by which this can occur in a neural field.

In the case of a Heaviside firing rate function (\ref{H}), cosine synaptic weight (\ref{cos}), and cosine input (\ref{cosn}), we have that the diffusion coefficient $D( \ve )$ will be given by the formula (\ref{badddeH}) and the mean reversion rate (\ref{kapinp}) will be given by
\begin{align}
\kappa  = \frac{n I_0 \sin (na)}{2 \sin^2 a + n I_0 \sin (na)}.   \label{coskap}
\end{align}
Not surprisingly, up to a scaling factor, this is the same as the eigenvalue (\ref{inpoddev}) associated with linear stability of odd perturbations to the bump in the deterministic system. With the formula for $\kappa$ in hand, we can approximate the variance of the stochastic process $\Delta (t)$ by the formula (\ref{ouinput}). We compare this theory to an average across realizations in Fig. \ref{inppinfig} for the cases $n=1$ and $n=2$, showing it captures the saturating nature of the variance.

\subsection{Noise-induced switching between two attractors}

\begin{figure}
\begin{center} \includegraphics[width=12.5cm]{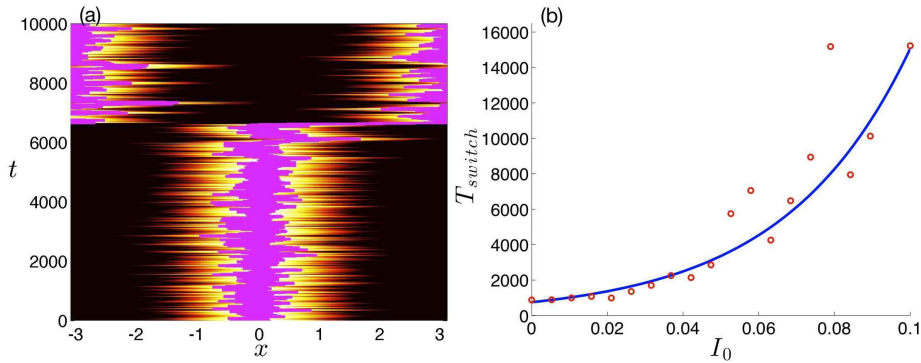} \end{center}
\caption{Escape of a pinned bump solution from the vicinity of one stable equilibrium to another. (a) Numerical simulation of the stochastic neural field (\ref{ringlang}) in the case $I(x) = I_0 \cos 2x$.  After a waiting time, the bump hops from $x \approx 0$ to $x \approx \pi$, the two stable fixed points of the underlying deterministic system. (b) Mean waiting time to a switch as a function of the strength of the input $I_0$ to the network as computed using numerical simulations (red circles). This is fit using least squares to an exponential $b \exp ( \gamma I_0 )$ (blue solid) where $b = 750$ and $\gamma = 30$. Other parameters are $\theta = 0.5$ and $\ve = 0.01$.}
\label{inpescfig}
\end{figure}

On substantially long waiting times, we would not necessarily expect $\Delta (t)$ to stay close to a fixed point of the deterministic system, even though we have made this assumption in our perturbation analysis. The bump will eventually escape to a neighboring fixed point (see Fig, \ref{inpescfig}(a)). Analogous to this, studies of mutually inhibitory neural networks have shown that including additive noise can cause transitions between two winner-take-all states of a network \cite{morenobote07}. To our knowledge, this is the first study to examine such phenomena in the context of a spatially extended neural field equation. However, there have been a studies of the switching times between wave propagation directions in a neural field with local adaptation that employed numerically derived forms of an effective potential \cite{laing01b,laing07}.

We find that additive noise causes trajectories of $\U (x,t)$ to cross through a separatrix of the deterministic system. Similar to our study of extinction in the input-free network, this separatrix is an unstable bump. Rather than separating a stable bump from a homogeneous ``off" state, here it separates two stable bumps from one another, centered at $x=0$ and $x= \pi$. In Fig. \ref{inpescfig}(a), we show one such transition. In this case, our approximation using an Ornstein-Uhlenbeck process (\ref{binpnsde}) will clearly break down, since the bump is now attracted to a completely different stable state. In Fig. \ref{inpescfig}(b), we show the the mean time until a switch $T_{switch}$ depends exponentially on the strength of the input $I_0$, given $b \exp ( \gamma I_0 )$. Essentially, we are controlling the depth of a bistable potential well in which the dynamics of the bump's position will evolve. The stronger the input, the deeper the well will be. As in the case of bump extinction, we might expect a Kramer's escape rate calculation could give us such a result \cite{risken84,gardiner04}. However, we will leave such calculations to future studies of rare events in neural fields.

\section{Pinning and reduced diffusion due to synaptic heterogeneity}
\label{inhom}

Synaptic connectivity that is patchy and periodic has been identified in anatomical studies of prefrontal cortex \cite{levitt93} and visual cortex \cite{angelucci02} using fluorescent tracers. Motivated by these findings, several mathematical analyses of stationary bumps in neural fields have employed weight kernels with periodic spatial heterogeneities \cite{laing02,laing03}. They found that such heterogeneities can lead to multiple bump solutions, where several disjoint subdomains of the network are active. In addition, some studies have examined the effects that synaptic weight heterogeneities have upon the propagation of traveling waves \cite{bressloff01,kilpatrick08,coombes11}, showing they can slow waves or even cause failure.

We explore the effect synaptic heterogeneities have on the diffusion of bumps. Noise causes bumps to wander freely in the translationally symmetric network, so the memory of the initial condition deteriorates over time. However, previous studies of bumps in spiking networks with some spatially dependent heterogeneity in model parameters have shown the bump will become pinned to a few discrete positions in the network \cite{zhang96,renart03}. Such symmetry breaks in the synaptic landscape of a network could originate from Hebbian plasticity reinforcing regions that have received more input during, for example, short term memory task training \cite{durstewitz00}. Mathematically, this can be understood as the dynamic landscape of the network switching from a line attractor to a chain of discrete attractors, just as we found in the input-driven network (\ref{inplang}). Here, we study a periodic heterogeneity in synaptic weights, which allows us to predict the most likely position for bumps to be. Interestingly, as the frequency of this heterogeneity is increased, so too does the effective diffusion of the bump.

\subsection{Existence of bumps}
We first show that the network (\ref{ring}) with a modified weight kernel (\ref{cosinh}) supports stationary bump solutions. While multibump solutions may arise in networks with certain periodic synaptic heterogeneities \cite{laing02}, only single bumps arise in the network with the weight kernel (\ref{cosinh}). This can be easily shown by applying the identity (\ref{cosid}) to the stationary solution problem ($u(x,t) = U(x)$). We will show that there are $2n$ locations $x = m \pi / n$ ($m \in \{ -n, ..., n-1 \}$) at which bumps can reside, rather than a continuum (centered at $x \in [- \pi , \pi ]$), as in the network with a translationally symmetric kernel like (\ref{cos}).

To start, we show the breakdown in the translation symmetry of stationary bump solutions in a network with general firing rate function $f$. Looking for stationary solutions $u(x,t) = U(x)$, we find that (\ref{ring}) with the periodic heterogeneous weight kernel (\ref{cosinh}) becomes
\begin{align}
U(x) = \int_{-\pi}^{\pi} (1 + \sigma w_1(n y)) \bar{w} (x-y) f(U(y)) \d y,  \label{bheter}
\end{align}
where $w_1(x) = w_1(x+2 \pi)$. Applying an arbitrary translation $b$ to the spatial argument of the bump, we find
\begin{align*}
U(x+b) &= \int_{- \pi}^{\pi} (1 + \sigma w_1(ny)) \bar{w} (x -y) f(U(y+b)) \d y \\
& = \int_{- \pi}^{\pi} (1 + \sigma w_1(nz-nb)) \bar{w} ( x + b - z) f(U(z)) \d z.
\end{align*}
Thus, if $\sigma >0$, then we must restrict $b = 2 \pi m/n$ for any $m \in \Z$, so that $w_1(nz-nb) = w_1(nz- 2\pi m) = w_1(z)$. Therefore, if $U(x)$ is a stationary bump solution to (\ref{ring}) with weight (\ref{cosinh}), then $U(x + 2 \pi m/n)$, with $m \in \Z$, is also a solution. Note also that reflection symmetry is preserved since
\begin{align*}
U(-x) &= \int_{- \pi}^{\pi} (1 + \sigma w_1 (ny)) \bar{w} ( x - y) f(U(-y)) \d y \\
&= \int_{- \pi}^{\pi} ( 1 + \sigma w_1 (-nz)) \bar{w} ( x + z ) f(U(z)) \d z \\
&= \int_{- \pi}^{\pi} ( 1 + \sigma w_1 (nz)) \bar{w} ( -x - z) f(U(z)) \d z,
\end{align*}
where we have used the facts that $w_1$ and $\bar{w}$ are even functions.

\begin{figure}
\begin{center} \includegraphics[width=12.5cm]{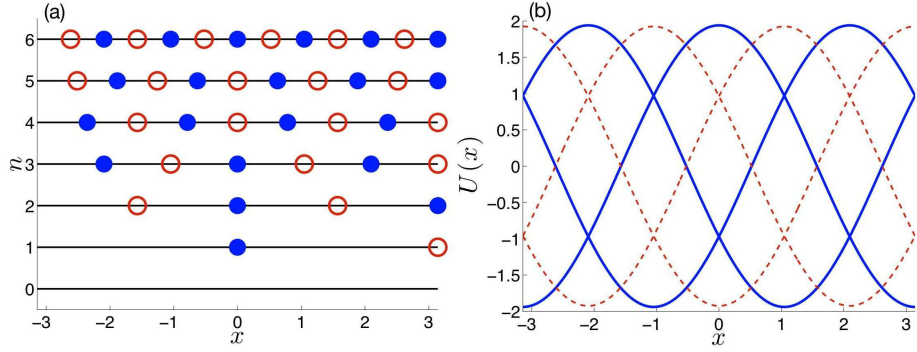} \end{center}
\caption{A finite number ($2n$) of bump locations in the network (\ref{ring}) having heterogeneous synaptic connectivity (\ref{cosinh}) with modulation frequency $n$. (a) Bump center locations along $x \in (- \pi , \pi]$ for various values of $n$ have an alternating pattern of locations with a stable bump (blue filled) and only unstable bumps (red circles). This creates a dynamic landscape of alternating stable nodes and saddles in space. (b) Associated bumps determined by implicit equation (\ref{inhbwid}) when $n=3$. Stable bumps with amplitude $A_+$ (\ref{inhApamp}) centered at $x=0, \pm \frac{2 \pi}{3}$ (blue solid). Unstable bump with amplitude $A_-$ (\ref{inhAmamp}) centered at $x = \pi, \pm \frac{\pi}{3}$ (red dashed). There are six other unstable bumps (not shown) that accompany each displayed bump. Other parameters are $\theta = 0.5$ and $\sigma = 0.2$. Firing rate function is Heaviside (\ref{H}).}
\label{binhcenters}
\end{figure}

Not only will bumps defined by (\ref{bheter}) exist in the network (\ref{ring}) with heterogeneous weight (\ref{cosinh}), there will also be bumps centered at $x=(2m + 1) \pi / n$, $m \in \Z$. However, these will have different amplitude than those centered at $x=2m \pi / n$. Upon writing
\begin{align*}
U(x + \pi / n) = \int_{- \pi}^{\pi} ( 1 + \sigma w_1 ( n y ) \bar{w}(x  - y) f(U(y + \pi /n)) \d y,
\end{align*}
a change of variables yields
\begin{align}
U(x + \pi / n) = \int_{ - \pi}^{\pi} (1 + \sigma w_1 ( nz - \pi ) ) \bar{w} ( x + \pi/ n - z) f(U(z)) \d z,  \label{Uinhoff}
\end{align}
so that for the weight function (\ref{cosinh}), equation (\ref{Uinhoff}) will become
\begin{align*}
U(x + \pi / n) = \int_{- \pi}^{\pi} ( 1 - \sigma w_1 ( ny ) ) \bar{w} (x + \pi / n - y) f(U(y)) \d y,
\end{align*}
where we have used the fact that $w_1( n x - \pi ) = \cos (nx - \pi ) = - \cos (nx)$. Using the same arguments as for bumps centered at $x=2 \pi m /n$, there will also be bumps centered at $x=(2m + 1) \pi / n$. Therefore, there will be $2n$ total bump locations in the network. We can use equation (\ref{bheter}) along with the weight function (\ref{cosinh}) to provide an amplitude equation for the bump $U = A_+ \cos x$ centered at $x=0$
\begin{align}
A_+ = \int_{- \pi}^{\pi} \cos x (1 + \sigma \cos (nx)) f(U(x)) \d x.  \label{inhApamp}
\end{align}
Similarly, the bump $U = A_- \cos(x + \pi /n)$, centered at $x= \pi / n$, will have amplitude
\begin{align}
A_- = \int_{ - \pi}^{\pi} \cos ( 1 - \sigma \cos (nx)) f(U(x)) \d x.  \label{inhAmamp}
\end{align}
We demonstrate how the number and stability of bumps depends on $n$ by plotting the bump centers on the domain $x \in [ - \pi , \pi ]$ for various values of $n$ in Fig. \ref{binhcenters}(a). Notice that as $n$ is increased, the $x=0$ bump reverses its stability at particular values of $n$. This result will be computed in our analysis of linear stability.

For a more illustrative analysis, we study the case of a Heaviside firing rate function (\ref{H}). Under this assumption, we can state the problem of looking for bump solutions $u(x,t) = U(x)$ by giving the requirement $U(x)> \theta$ for $x \in ( - a , a)$ and $U(x)< \theta$ otherwise so we can compute the amplitude
\begin{align}
A_+ = \int_{-a}^{a} \cos x + \sigma \cos x \cos n x \d x,   \label{Ainh}
\end{align}
and the other amplitude is given
\begin{align}
A_- = \int_{-a}^{a} \cos x - \sigma \cos x \cos n x \d x.  \label{Asinh}
\end{align}
Thus, we can see that switching the sign of $\sigma$ will still yield the same set of bump solutions, but they will be centered at different places.

\begin{figure}
\begin{center}  \includegraphics[width=12.5cm]{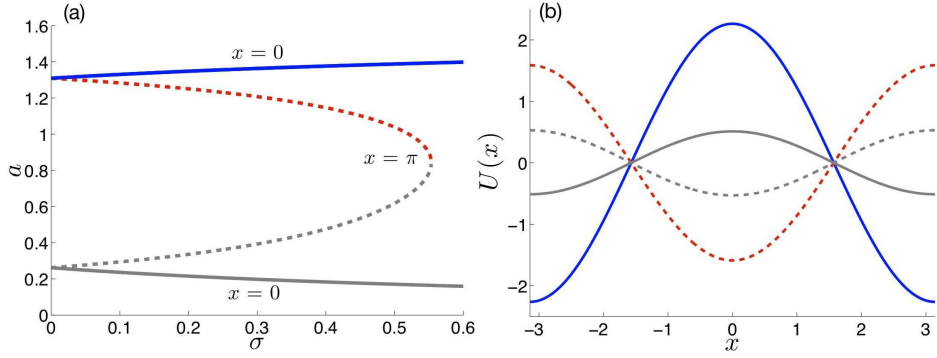} \end{center}
\caption{Bumps in the ring network (\ref{ring}) using heterogeneous synaptic connectivity (\ref{cosinh}) with modulation frequency $n=1$. (a) Bump half-width $a$ as it depends on amplitude of heterogeneity amplitude $\sigma$. A wide bump centered at $x= \pi$ (red dashed) separates the wide stable bump at $x=0$ (blue solid) from itself on the periodic domain. Narrow bump at $x=0$ (grey solid) separates stable bump from homogeneous off state. (b) Profile of each bump for $\sigma = 0.2$. Threshold parameter is $\theta = 0.5$. Firing rate function is Heaviside (\ref{H}).}
\label{inhn1bumps}
\end{figure}

Now, we compute the bump amplitudes (\ref{Ainh}) and (\ref{Asinh}), which only differ in the sign of $\sigma$. First, we analyze the special case $n=1$, in which $A_{\pm}$ can be integrated
\begin{align*}
A_{\pm} = 2 \sin a \pm \sigma a \pm \frac{\sigma}{2} \sin (2a).
\end{align*}
Invoking the threshold condition $U( \pm a) = \theta$, we can generate and implicit equation for the bump half-width $a$ given by
\begin{align}
\theta = \sin 2a \pm \sigma \left[ a \cos a + \frac{\sin a + \sin (3a)}{4} \right].  \label{inhn1bwid}
\end{align}
Per our general analysis of the symmetry of bump solutions, we expect there to only be one peak location for each sign of $\sigma$ ($x=0$ and $x= \pi$), since the period of $w_1$ in this case is $2\pi$, the length of the domain. However, as in the case of the homogeneous weight function, there can be two half-widths $a$ at each location. As we can compute using linear stability, a maximum of one bump at each position of these will be linearly stable. This is demonstrated in Fig. \ref{inhn1bumps}.
 
In the case that $n>1$, we can integrate (\ref{Ainh}) so that we find
\begin{align}
A_{\pm} = 2 \sin a \pm \sigma \left[ \frac{\sin ((n-1)a)}{n-1} + \frac{\sin ((n+1)a)}{n+1} \right].  \label{binhampl}
\end{align}
Upon requiring the threshold crossing conditions $U( \pm a ) = A_{\pm} \cos a = \theta$, we can implicitly specify the bump half-width with the equation
\begin{align}
\theta = \sin (2a) \pm \frac{\sigma}{2} \left[ \frac{\sin ((n-2)a)}{n-1} + \frac{2n \sin (na)}{n^2-1} + \frac{\sin ((n+2)a)}{n+1} \right].  \label{inhbwid}
\end{align}
Since $\cos x$ is a unimodal function, its sole maximum will occur at $x=0$ ($x = \pi / n$), when $A_+>0$ ($A_- > 0$). Therefore, we do not expect the appearance of multibump solutions in this context. We would only expect this if the heterogeneity in (\ref{cosinh}) were in the $x$ variable. We now proceed to study the linear stability of the bump solutions specified by (\ref{inhn1bwid}) and (\ref{inhbwid}).

\subsection{Stability of bumps}

We now study the stability of bumps in the network (\ref{ring}) with heterogeneous synaptic weights. As we observed in our existence analysis, switching the sign of $\sigma$ will lead to the two classes of bumps changing places. Therefore, we only study the stability of bumps centered at $x=0$, as simply flipping the sign of $\sigma$ will provide us with stability of the complementary bump. To analyze the stability of the bump, we study the evolution of small, smooth, separable perturbations to the bump $u(x,t) = U(x) + \psi (x) \e^{\lambda t}$, where $|\psi (x) | \ll 1$. Plugging into (\ref{ring}) and truncating to first order
\begin{align}
(\lambda + 1) \psi (x) = \int_{- \pi}^{\pi} w(x,y) f'(U(y)) \psi (y) \d y.   \label{stabinh}
\end{align}
The essential spectrum $\lambda = -1$ does not contribute to any instabilities here. Therefore, any instabilities can be identified by studying the point spectrum. We can identify these by appropriately manipulating the integral term in (\ref{stabinh}). In the case of the particular weight function (\ref{cosinh}), we can apply the identity (\ref{cosid}) to write the equation (\ref{stabinh}) as a $2 \times 2$ linear spectral problem
\begin{align*}
( \lambda + 1) \left( \begin{array}{c} {\mc A} \\ {\mc B} \end{array} \right) = \left( \begin{array}{cc} {\mc J}_n ( \cos^2 x ) & {\mc J}_n ( \cos x \sin x ) \\ {\mc J}_n ( \cos x \sin x ) & {\mc J}_n ( \sin^2 x ) \end{array} \right) \left( \begin{array}{c} {\mc A} \\ {\mc B} \end{array} \right)
\end{align*}
where
\begin{align*}
{\mc J}_n( r(x)) = \int_{- \pi}^{\pi} r (x) (1 + \sigma \cos (nx)) f'(U(x)) \d x.
\end{align*}
It is clear that, since $U(x)$ is even, ${\mc J}_n( \cos x \sin x)=0$. Therefore, the two eigenvalues describing the linear stability of the bump (\ref{bheter}) will be one associated with odd perturbations
\begin{align*}
\lambda_o = - 1 + {\mc J}_n( \sin^2 x)
\end{align*}
and one associated with even perturbations
\begin{align*}
\lambda_e = - 1 + {\mc J}_n( \cos^2 x).
\end{align*}

\begin{figure}
\begin{center} \includegraphics[width=12.5cm]{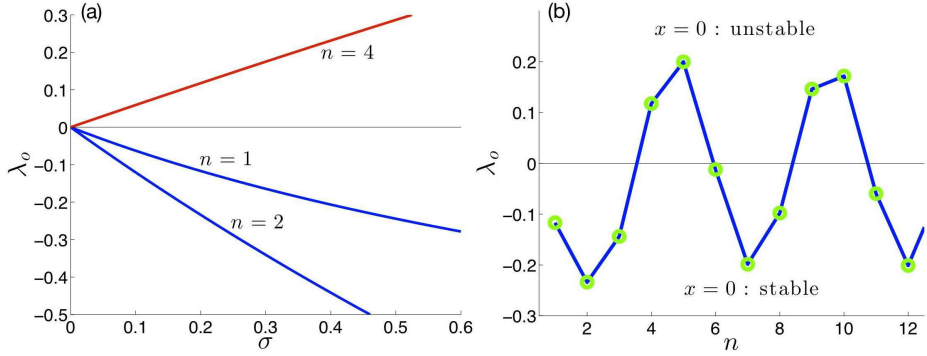} \end{center}
\caption{Eigenvalue $\lambda_o$ associated with odd perturbations of the bump centered at $x=0$ given by (\ref{bheter}) with amplitude $A_+$ specified (\ref{Ainh}). (a) Eigenvalue $\lambda_-$ as a function of heterogeneity amplitude $\sigma$ becomes negative, indicating linear stability, when $n=1$ and $n=2$ but become positive, indicating linear instability, when $n=4$. (b) Eigenvalue $\lambda_o$ as a function of synaptic modulation frequency $n$ as determined by the formulae (\ref{evoddn1}) for $n=1$ and (\ref{evoddng1}) for $n>1$. Heterogeneity amplitude is fixed $\sigma = 0.2$. Threshold parameter $\theta = 0.5$.}
\label{bheteroeval}
\end{figure}

In the case of a Heaviside firing rate function (\ref{H}), we can use (\ref{dH}) to calculate the integral terms
\begin{align*}
{\mc J}_n ( \sin^2 x) &= \frac{2 \sin^2 a + 2 \sigma \cos (n a) \sin^2 a}{A_+ \sin a} \\
{\mc J}_n ( \cos^2 x) & = \frac{2 \cos^2 a + 2 \sigma \cos (na) \cos^2 a}{A_+ \sin a}.
\end{align*}
To study the effect of heterogeneities on the eigenvalues, we start with the special case $n=1$. Here, the eigenvalue associated with odd perturbations is given
\begin{align}
\lambda_o = -1 + \frac{2 \sin a + 2 \sigma \sin a \cos a}{2 \sin a + \sigma a + 2 \sigma \sin a \cos a} = - \frac{\sigma a}{2 \sin a + \sigma a + \sigma \sin 2 a} < 0,  \label{evoddn1}
\end{align}
since $a>0$. Thus, we can be certain that the bump is linearly stable to shift perturbations when $n=1$ and $\sigma > 0$. In a complementary way, bumps in the network where $\sigma < 0$ will be linearly unstable to shift perturbations when $n=1$. Even perturbations have associated eigenvalue
\begin{align*}
\lambda_e = -1 + \cot^2 a \frac{2 \sin a + 2 \sigma \sin a \cos a}{2 \sin a + \sigma a + 2 \sigma \sin a \cos a}. 
\end{align*}
For $n>1$, the eigenvalue associated with odd perturbations will be
\begin{align}
\lambda_o &= -1 + \frac{1 + \sigma \cos n a}{1 + \frac{\sigma}{n^2 -1} \left[ n \cot a \sin ( n a) - \cos ( n a) \right]}, \\
&= \frac{\sigma n [ n \sin a \cos (na) - \cos a \sin (na)]}{(n^2-1) \sin a + \sigma [ n \cos a \sin (na) - \sin a \cos (na)]}  \label{evoddng1}
\end{align}
which will, in general, not be zero. We plot the eigenvalue $\lambda_o$ as a function of $\sigma$ and of $n$ in Fig. \ref{bheteroeval}. As we have mentioned, the eigenvalue $\lambda_o$ oscillates as a function of $n$ so that the bump at $x=0$ reverses its stability. The eigenvalue associated with even perturbations of the bump will be given by
\begin{align*}
\lambda_e = -1 + \cot^2 a \frac{1 + \sigma \cos n a}{1 + \frac{\sigma}{n^2 -1} \left[ n \cot a \sin ( n a) - \cos ( n a) \right]}.
\end{align*}
In general, its sign will not change for small amplitudes $\sigma$.

\subsection{Pinned bumps for low frequency modulation}

To analyze the effect that noise has upon bump solutions, we will begin by making a small noise assumption and performing an asymptotic expansion, as we did for the homogeneous network. Due to spatial heterogeneities, noise causes the center of the bump to move as a mean-reverting stochastic process, rather than a purely diffusive process. Synaptic heterogeneities, however subtle, can trap neural activity in basins of attraction whose widths are defined by the period of the heterogeneity (\ref{cosinh}). On exponentially long timescales we would expect the bump to escape from these potential wells. However, even for weak heterogeneities, escape rates are low enough such that the movement of the bump away from its initial condition can occur more slowly than in the homogeneous case.

Our analysis here is mainly concerned with the effect periodic heterogeneities have upon the diffusion of bumps. Thus, we merely consider the case of additive noise ($g(\U) = 1$), so the noise will have zero mean. Though the case of multiplicative noise could be analyzed, that of additive noise makes for more transparent results. We assume the additive noise in (\ref{ringlang}) generates two phenomena that occur on distinct timescales. The center of the bump will fluctuate about its original position on long time scale according to the stochastic variable $\Delta (t)$. On short timescales, the profile of the bump $U$ will fluctuate according to the series of higher order corrections $\ve^{1/2} \Phi + \ve \Phi_1 + \ve^{3/2} \Phi_2 + \cdots$. Thus, we plug the expansion (\ref{binpexp}) into (\ref{ringlang}) and study the hierarchy of equations generated by expanding in powers of $\ve^{1/2}$. To leading order, we find the deterministic equation (\ref{bheter}) for the mean bump profile $U(x)$. To next order, we find that $\Delta (t) = \Oo ( \ve^{1/2})$ and
\begin{align}
\d \Phi (x,t) = \Lo \Phi (x,t) \d t + \ve^{-1/2} U'(x) \d \Delta (t) + \d W( x, t) + \ve^{-1/2} B(x) \Delta (t) \d t   \label{dpinhom}
\end{align}
where $\Lo$ is the non-self-adjoint linear operator (\ref{linopad}), and
\begin{align}
B(x) = \sigma n \int_{- \pi}^{\pi} w_1'(ny) \bar{w} ( x- y) f(U(y)) \d y.  \label{bmBinh}
\end{align}
The last term on the right hand side of (\ref{dpinhom}) is generated by integrating the heterogeneous contribution from the weight function (\ref{cosinh}) by parts and truncating with a linearization. Notice that since $B(x)$ scales with $n$, this approximation will only be valid for small enough $n$ values. Thus, we only consider the effect of low modulation frequencies $n$ in this subsection. Now, we can ensure that a bounded solution to (\ref{dpinhom}) exists by requiring the inhomogeneous part is orthogonal to the nullspace $\vp (x)$ of the adjoint operator $\Lo^*$ defined by (\ref{adjopad}). Upon taking the $L^2$ inner product of both sides of (\ref{dpinhom}) with $\vp (x)$, we have the solvability condition
\begin{align}
\int_{- \pi}^{\pi} \vp ( x) \left[ U'(x) \d \Delta (t) + B(x) \Delta (t) \d t + \ve^{1/2} \d W(x, t) \right] \d x = 0.   \label{sdeinhom}
\end{align}
We can then rearrange the stochastic differential equation (\ref{sdeinhom}) to find that $\Delta (t)$ satisfies the Ornstein-Uhlenbeck process
\begin{align}
\d \Delta (t) + \kappa \Delta (t) \d t = \d {\mc W} (t),    \label{ouinhom}
\end{align}
where
\begin{align}
\kappa = \frac{\int_{- \pi}^{\pi} \vp ( x ) B(x) \d x}{\int_{- \pi}^{\pi} \vp (x) U'(x) \d x}  \label{kapheter}
\end{align}
and ${\mc W}(t)$ is the white noise process defined by (\ref{whiteinp}) having diffusion coefficient $D( \ve )$ (\ref{bdiffcoeff}) with $g( \U ) = 1$, as in the case of the input-driven network (\ref{inplang}). Assuming we start the bump upon a stable attractor, as defined by our existence and stability calculations of (\ref{ring}) with synaptic weight (\ref{cosinh}), we can calculate the mean and variance of the Ornstein-Uhlenbeck process (\ref{ouinhom}) using standard techniques \cite{gardiner04} to find
\begin{align}
\langle \Delta (t) \rangle = 0, \ \ \ \ \ \langle \Delta (t)^2 \rangle - \langle \Delta (t) \rangle^2 = \frac{D( \ve )}{2 \kappa} \left[ 1 - \e^{-2 \kappa t} \right]. \label{oumv}
\end{align}
This provides us with a different result than the translationally symmetric system with $w(x,y) = \bar{w}(x-y)$ where the bump freely diffuses. However, it is related to the result we found in our input driven system (\ref{inplang}) where an external input (\ref{cosn}) of frequency $n$ locks the bump to the vicinity of a $n$ discrete attractors. Here the bump is pinned by internal bias generated by the heterogeneous contribution of the weight kernel $w_1(ny)$, so that the variance saturates at $D( \ve )/ \kappa$ for large values of $t$, according to the approximation (\ref{oumv}).

\begin{figure}
\begin{center} \includegraphics[width=12.5cm]{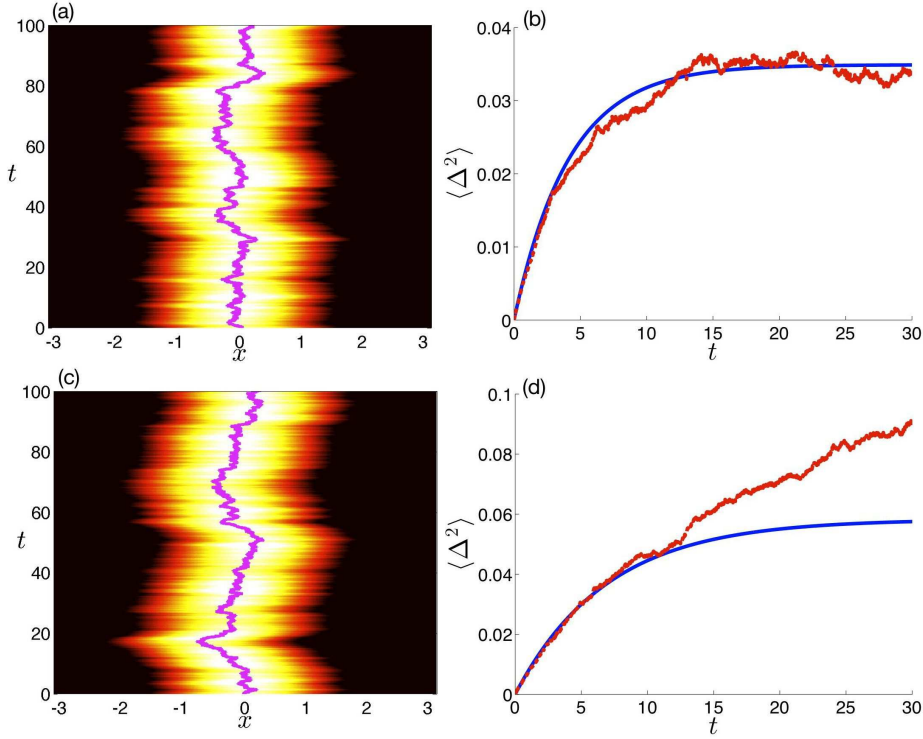} \end{center}
\caption{Pinning of bumps in the network (\ref{ringlang}) with synaptic weight (\ref{cosinh}) for low frequency $n$ synaptic heterogeneity. (a) Numerical simulation of (\ref{ringlang}) using synaptic weight (\ref{cosinh}) for $n=2$, $\sigma = 0.1$, and $\ve = 0.01$ shows bump remains pinned to the stable attractor at $x=0$. (b) Variance of the bump's position plotted against time computed numerically (red dashed) across 1000 realizations saturates after a moderate amount of time when $n=2$, as predicted by the Ornstein-Uhlenbeck approximation (\ref{ouinhom}) (blue solid). (c) Numerical simulation for $n=3$, $\sigma = 0.1$, and $\ve = 0.01$ shows bump remains pinned to the stable location at $x=0$. (d) Variance of the bump's position plotted against time computed numerically (red dashed) does not match the prediction of the Ornstein-Uhlenbeck approximation (blue solid) quite as well for long times. Threshold parameter $\theta = 0.5$}
\label{inhlosat}
\end{figure}

In order to study our asymptotic analysis, we consider the case of a Heaviside firing rate function (\ref{H}), cosine (\ref{cos}) for $\bar{w}$ and $w_1$, and cosine spatial noise correlations (\ref{coscorr}). In this case, the diffusion coefficient $D(\ve)$ is given by the formula (\ref{badddeH}). In addition, we restrict our modulation frequency to be greater than unity, $n>1$. Then the function $B(x)$, which leads to pinning, can be computed, using the formula (\ref{vpadsol}) for $\vp (x)$ so
\begin{align*}
B(x) = -\sigma n \int_{-a}^{a} \sin (n y) \cos ( x- y) \d y = - 2 \sigma n \frac{\cos a \sin (na) - n \sin a \cos (na)}{n^2-1} \sin x.
\end{align*}
Plugging this into our formula for the mean reversion rate (\ref{kapheter}), where we use our formula for the amplitude of the bump (\ref{binhampl}), we have
\begin{align}
\kappa = \frac{\sigma n [ \cos a \sin (na) - n \sin a \cos (na) ] }{(n^2-1) \sin a \pm \sigma [n \cos a \sin (na) - \sin a \cos (na) ]},  \label{kapcos}
\end{align}
which, not surprising, is simply the eigenvalue $\lambda_-$ associated with odd perturbations (\ref{evoddng1}), up to a sign switch. The sign of the $\sigma$ portion of the denominator is ambiguous because we must select the stable bump, which could have either $A_+$ or $A_-$ as its amplitude. Using these specific formulae, we can compute the variance of the Ornstein-Uhlenbeck process (\ref{ouinhom}) with the formula (\ref{oumv}). We show example of this in Fig. \ref{inhlosat} for $n=2$ and $n=3$. In particular, we observe that the variance of the bump, computed by averaging across many realizations of (\ref{ringlang}) saturates after a substantial amount of time. However, as the number of attractors is increase, the Ornstein-Uhlenbeck approximation (\ref{ouinhom}) does not do as well approximating the variance, since the bump can begin to escape from the starting pinned location to a neighboring one.

\subsection{Reduced diffusion of bumps for high frequency modulation}

\begin{figure}
\begin{center} \includegraphics[width=12.5cm]{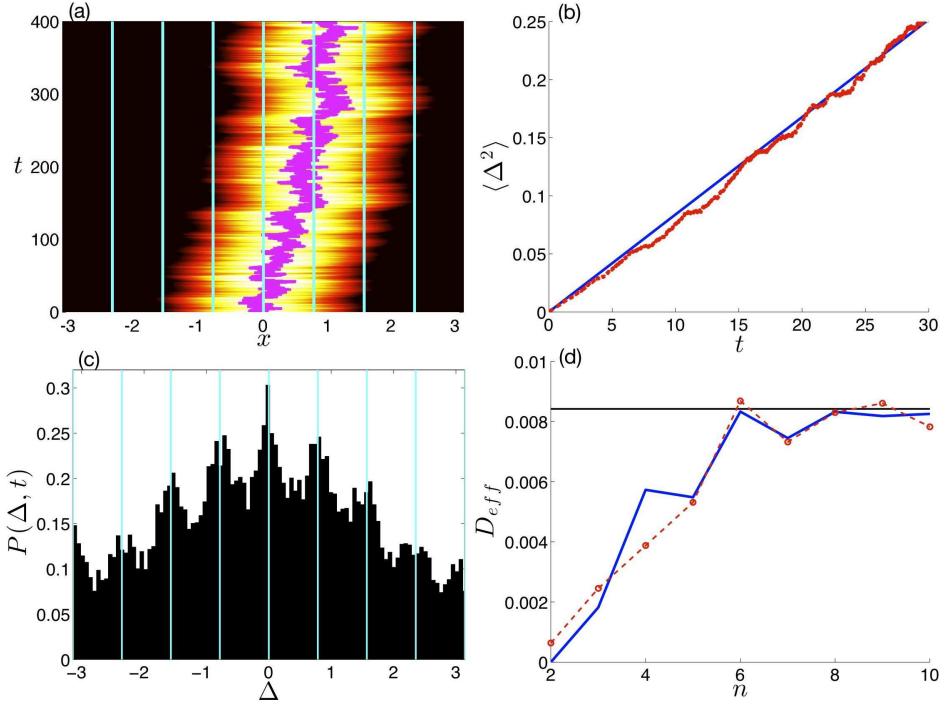} \end{center}
\caption{Reduced effective diffusion in a network with high frequency modulation in synaptic weights. (a) Numerical simulation of (\ref{ringlang}) with synaptic weight (\ref{cosinh}) in the case $n=8$ where the bump makes frequent jumps between locations of stable attractors (cyan) of the deterministic system. (b) Variance in the bumps position scales linearly with time, rather than saturating as in the case of lower frequency modulation of synaptic weights. (c) Probability density $P( \Delta , t)$ of bump position computed across 5000 realizations evaluated at time $t=400$ reveals microperiodic structure of diffusion suggested by (\ref{pergauss}). Vertical lines (cyan) indicate the location of the $n=8$ attractors. (d) Asymptotic approximation of effective diffusion $D_{eff}$ (blue line) computed using theory (\ref{deffcos}) as compared with that computed using numerical simulations (red dashed dot). For small values of $n$, effective diffusion is considerably reduced as compared to diffusion (\ref{badddeH}) in the homogeneous system (black line). Other parameters are $\theta = 0.5$, $\sigma = 0.1$, and $\ve = 0.01$.}
\label{inhomhifreq}
\end{figure}

As opposed to the expansion we performed in the previous subsection, we could consider a perturbative approximation that takes into account the nonlinearity of the synaptic heterogeneity (\ref{cosinh}), rather than linearizing it to yield the Ornstein-Uhlenbeck approximation (\ref{ouinhom}). To do so, we can take note of the fact that, as $n$ becomes large, the contribution made by the heterogeneous part of (\ref{cosinh}) becomes small. Thus, it is not necessary to perform an expansion of this portion in $\Delta (t)$ in order to truncate the integral term in (\ref{ringlang}). In fact, doing so would cause ever worse approximation, due to the slope of the linearization (\ref{bmBinh}) becoming steeper and steeper, as it scales with $n$. This is related to the fact that as $n$ increases, the bump begins to escape from the vicinity of individual discrete attractors more often (see Fig. \ref{inhomhifreq}). As before, we can perform the expansion (\ref{binpexp}) where $U$ is a stationary bump solution (\ref{bheter}), $\Delta (t)$ tracks the wandering of the bump, and $\Phi$ tracks fast fluctuations in the profile of the bump. Plugging this into (\ref{ringlang}), we perform a similar averaging and hierarchy expansion to before. The ${\mc O}( \ve^{1/2})$ equation is then given
\begin{align}
\d \Phi (x,t) = {\mc L} \Phi (x,t) \d t + \ve^{-1/2} U'(x) \d \Delta (t) + \d W(x,t) + \ve^{-1/2} B(x, \Delta (t)) \d t,  \label{dphihfreq}
\end{align}
where ${\mc L}$ is the non-self-adjoint linear operator (\ref{linopad}), and
\begin{align}
B(x, \Delta ) = \sigma \int_{ - \pi }^{\pi} [ w_1 (n(y + \Delta )) - w_1 (n y) ] \bar{w}  ( x - \Delta - y) f(U(y)) \d y,  \label{Bnonlin1}
\end{align}
which we will show to be small below. To derive the function $B(x, \Delta )$, we have performed the change of variables
\begin{align*}
\int_{- \pi}^{\pi} w(x,y) f(U( y - \Delta)) \d y &= \int_{ - \pi}^{\pi} w (x, z + \Delta)  f(U ( z)) \d z \\
&= \int_{- \pi}^{\pi} (1 + \sigma w_1 (ny)) \bar{w} (x - \Delta - y ) f(U (y)) \d y \\
 & + \sigma \int_{- \pi}^{\pi} ( w_1 ( n ( y + \Delta ) ) - w_1(ny)) \bar{w} (x - \Delta - y) f( U (y)) \d y, 
\end{align*}
in order to make the cancellation
\begin{align*}
U(x- \Delta ) = \int_{- \pi}^{\pi} (1 + \sigma w_1 (ny) ) \bar{w} (x - \Delta -y) f(U(y)) \d y.
\end{align*}
Since $w_1 (ny)$ is a $2 \pi /n$--periodic function, we can also assume that $B(x, \Delta )$ will be $2 \pi / n$--periodic in $ \Delta$. To justify the retention of the term $B(x, \Delta )$ in the ${\mc O}( \ve^{1/2})$, we note that upon integrating (\ref{Bnonlin1}) by parts, we have
\begin{align*}
B(x, \Delta ) = \frac{\sigma}{n} \int_{- \pi}^{\pi} W_d (ny) \frac{\d}{\d y} \left[ \bar{w}(x - \Delta - y) f(U(y)) \right] \d y = {\mc O} (1/n),
\end{align*}
which will be small for large $n$, and we have defined
\begin{align*}
W_d (x) = \int_{- \pi}^x \left[ w_1(y+ \Delta ) - w_1 (y) \right] \d y.
\end{align*}
Note also that since we require $\Delta (t) = {\mc O} ( \ve^{1/2} )$ for our approximation, we can truncate a Taylor expansion of $\bar{w}(x - y - \Delta )$ so that we define
\begin{align}
B(x, \Delta) = \sigma \int_{ - \pi}^{\pi} \left[ w_1 (n(y + \Delta )) - w_1 (ny) \right] \bar{w} (x - y) f(U(y)) \d y.  \label{Bnonlin}
\end{align}
Now, we can ensure that a bounded solution to (\ref{dphihfreq}) exists by requiring the inhomogeneous part is orthogonal to the nullspace $\vp (x)$ of the adjoint operator ${\mc L}^*$ defined by  (\ref{adjopad}). Upon taking the $L^2$ inner product of both sides of (\ref{dphihfreq}) with $\vp (x)$, we have the sufficient solvability condition
\begin{align}
\int_{- \pi}^{\pi} \vp (x) \left[ U'(x) \d \Delta (t) + B(x, \Delta ) \d t + \ve^{1/2} \d W(x,t) \right] \d x = 0. \label{nlnsde}
\end{align}
We can then rearrange the solvability condition (\ref{nlnsde}) to find that $\Delta (t)$ satisfies the nonlinear stochastic differential equation
\begin{align}
\d \Delta (t) + K (n \Delta ) \d t = \d {\mc W} (t),  \label{Knlnsde}
\end{align}
where
\begin{align}
K (n \Delta ) = \frac{\int_{- \pi}^{\pi} \vp ( x) B( x, \Delta) \d x }{\int_{- \pi}^{\pi} \vp ( x) U '(x) \d x}  \label{Knonlin}
\end{align}
is a $2 \pi /n$--periodic function since $B(x, \Delta)$ is $2 \pi  /n$--periodic in $\Delta$, and the white noise process ${\mc W} (t)$ still defined by (\ref{whiteinp}) having diffusion coefficient $D( \ve)$ given by (\ref{bdiffcoeff}) with $g( \U ) = 1$, as before. Therefore, we have reduced the problem of a bump wandering in a neural field with periodic synaptic microstructure to that of a particle diffusing in a periodic potential. This is a well studied problem for which it is possible to derive an effective diffusion coefficient \cite{risken84}. To do so, we must derive the profile of the periodic potential well governing the dynamics. To find this, we simply integrate the nonlinear function (\ref{Knonlin}), which yields
\begin{align}
V( \Delta ) = \int_{ - \pi}^{\Delta} K(n \eta ) \d \eta = \frac{\int_{- \pi}^{\pi} \vp (x) \int_{ - \pi}^{\Delta} B(x, \eta ) \d \eta \d x}{\int_{- \pi}^{\pi} \vp (x) U'(x) \d x}.  \label{perpwell}
\end{align}
With the $2 \pi/n$-periodic potential well (\ref{perpwell}) in hand, we can derive the effective diffusion coefficient
\begin{align}
D_{eff} = \lim_{t \to \infty} \frac{\langle \Delta(t)^2 \rangle - \langle \Delta (t) \rangle^2}{t}  \label{defft}
\end{align}
of the stochastic process defined by (\ref{Knlnsde}). As the definition of $D_{eff}$ (\ref{defft}) suggests, the approximation is valid in the limit of large time. However, we do find that it works quite well for reasonably short times too. This is contingent upon the modulation frequency $n$ being substantially large. As many authors have found, this approximation arises from the fact that the density of trajectories tends asymptotically to \cite{lifson62,lindner01}
\begin{align}
P_{as} ( \Delta ,t) = P_0( \Delta ) \frac{\exp [ - \Delta^2 / 4 D_{eff} t ]}{\sqrt{4 \pi D_{eff} t}}  \label{pergauss}
\end{align}
where $P_0$ refers to the stationary ($2 \pi / n$--periodic) solution of (\ref{Knlnsde}). This function is responsible for microstructure of the density whereas the Gaussian is responsible for it macrostructure. Usually, this structure is numerically extracted by evolving the Fokker-Planck formalism of the Langevin equation (\ref{Knlnsde}), so the approximation (\ref{pergauss}) can be made as an ansatz. In this case, we can approximate using the Lifson-Jackson formula \cite{lifson62,festa78,risken84}
\begin{align}
D_{eff} = \frac{ D ( \ve ) ( 2 \pi /n)^2}{ \int_0^{2 \pi / n} \int_0^{2 \pi / n} \exp \left[ \frac{2 (V(x) - V(y))}{D( \ve )} \right] \d y \d x} , \label{deffform}
\end{align}
where we have used the diffusion coefficient $D( \ve )$ of the white noise source and the $2 \pi / n$--periodicity of the potential well (\ref{perpwell}). As we will show, the heterogeneity introduced in the synaptic weight (\ref{cosinh}) tends to decrease the effective diffusion coefficient. In other words, we usually find that $D_{eff} < D ( \ve )$. Thus in some sense, having a chain of discrete attractors appears to provide better memory of the initial condition than a line attractor. Of course the trade off is that only a finite number of initial conditions, specifically $n$, can be represented in our network (\ref{ringlang}) with the weight (\ref{cosinh}) with modulation frequency $n$. We will explore this issue further in future studies.

For now, we compare the asymptotic approximation of $D_{eff}$ to numerical simulations. We thus consider the case of a Heaviside firing rate function (\ref{H}), cosine (\ref{cos}) for $\bar{w}$ and $w_1$, and cosine spatial correlations (\ref{coscorr}). In this case, the diffusion coefficient $D( \ve )$ is given by the formula (\ref{badddeH}). First, we compute the function $B(x , \Delta )$ (\ref{Bnonlin}), which is $ 2 \pi / n$--periodic in the $\Delta$ argument
\begin{align}
B( x , \Delta ) &= \sigma \left[ \frac{2 ( \cos (n \Delta ) - 1) ( n \cos a \sin (na) - \sin a \cos (na) )}{n^2 - 1} \right] \cos x \nonumber \\
& + \sigma \left[ \frac{2 \sin (n \Delta ) ( n \sin a \cos (na) - \cos a \sin (na ) )}{n^2 -1 } \right] \sin x.  \label{Bnlnfin}
\end{align}
Now, with the formula (\ref{Bnlnfin}) in hand, as well as (\ref{vpadsol}) for $\vp (x)$ and the equation for the amplitude $A_{\pm}$ (\ref{binhampl}), we can compute the nonlinear function $K(n \Delta )$ using (\ref{Knonlin}). Note that the cosine portion of (\ref{Bnlnfin}) vanishes upon integration to yield
\begin{align}
K( n \Delta ) = \left[  \frac{2 \sigma ( n \sin a \cos (na) - \cos a \sin (na) )}{(n^2-1) \sin a \pm \sigma [ n \cos a \sin (na) - \sin a \cos (na) ]} \right] \sin (n \Delta ),  \label{Kfin}
\end{align}
where we select the $+$ or $-$ in the denominator of (\ref{Kfin}), depending on whether the bump centered at $x=0$ is stable or not. Now, in order to compute our effective diffusion coefficient $D_{eff}$, we must integrate the function $K( n \Delta )$ to yield the potential function governing the dynamics of (\ref{Knlnsde}). This gives us the potential function
\begin{align}
V( \Delta ) &= - {\mc V} (n) \cos (n \Delta ), \label{Vcos}
\end{align}
where the amplitude (or half-height) of each well is
\begin{align}
{\mc V} (n) = \frac{2 \sigma (n \sin a \cos (na) - \cos a \sin (na))}{n (n^2 - 1) \sin a \pm \sigma n [ n \cos a \sin (na) - \sin a \cos (na)]}.  \label{ppamp}
\end{align}
Now, finally, we use the standard formula for the effective diffusion coefficient of a particle in a periodic potential well (\ref{deffform}). With our particular cosine potential well (\ref{Vcos}), we find that each integral can be computed and are equal
\begin{align*}
\int_0^{2 \pi /n} \exp \left[ \frac{2 V(x)}{D( \ve) } \right ] \d x &= \int_0^{2 \pi /n} \exp \left[ - \frac{V(x)}{D( \ve) } \right ] \d x \\
&= \int_0^{2 \pi /n} \exp \left[ \frac{2 {\mc V} (n)}{D( \ve) } \cos (nx) \right ] \d x = \frac{2 \pi}{n} I_0 \left( \frac{2 {\mc V} (n)}{D( \ve )} \right),
\end{align*}
where $I_0 (x)$ is the modified Bessel function of the zeroth kind. Therefore, the formula (\ref{deffform}) for the effective diffusion coefficient yields
\begin{align}
D_{eff} = \frac{D( \ve )}{[ I_0 ( 2 {\mc V} (n) / D ( \ve ) )]^2}.   \label{deffcos}
\end{align}
Using this formula along with the definition (\ref{ppamp}), we approximate the diffusion of a bump in a network with synaptic modulation frequency $n=8$ in Fig. \ref{inhomhifreq}(b). Notice, the linear approximation of the variance's scaling with time matches averages over realizations fairly well. Thus, the variance no longer saturates in time, as in the case of low frequency modulation $n$. As evidenced by our plots of the probability density $P( \Delta, t)$, in Fig. \ref{inhomhifreq}(c), the stochastic process $\Delta (t)$ behaves diffusively with microperiodic modulation, as suggested by the asymptotic formula (\ref{pergauss}). Now, we can note that in the limit of high amplitude modulations ($n \to \infty $) the formula (\ref{deffcos}) tends to the diffusion coefficient of the homogeneous network since
\begin{align*}
\lim_{n \to \infty} {\mc V}(n) = \lim_{n \to \infty} \frac{2 \sigma (n \sin a \cos (na) - \cos a \sin (na))}{n (n^2 - 1) \sin a \pm \sigma n [ n \cos a \sin (na) - \sin a \cos (na)]} = 0
\end{align*}
so that
\begin{align*}
\lim_{n \to \infty} I_0 \left( \frac{2 {\mc V}(n)}{D( \ve )} \right) &= I_0 (0) = 1,
\end{align*}
and thus we find the limit of (\ref{deffcos}) to be
\begin{align*}
\lim_{n \to \infty} D_{eff} = D ( \ve ).
\end{align*}
Since $I_0 (x)$ has a global minimum at $x=0$, it is clear that $D_{eff} < D ( \ve )$ for all $n$. However, in numerical simulations, we would also presume the effects of pinning, as described by (\ref{ouinhom}) would also be present. Nonetheless, we compare our theoretical effective diffusion (\ref{deffcos}) across a span of modulation frequencies $n$ to that approximated using numerical simulations in Fig. \ref{inhomhifreq} (d). We find reasonable agreement. In particular, we see the result that synaptic heterogeneity substantially reduces the effective diffusion of the bump for lower values of $n$. We plan to pursue this result much more deeply in future studies. 

\section{Discussion}
\label{disc}

We have analyzed the effects of external noise on stationary bumps in spatially extended neural field equations. In a network with spatially homogeneous synaptic weights, we found that noise causes bumps to wander about the spatial domain according to a purely diffusive process. We can asymptotically approximate the diffusion coefficient of this process using a small-noise expansion, which assumes the profile of the activity variable is still a bump to first order. Comparing the effects of purely additive and multiplicative noise, we find that multiplicative noise alters the mean amplitude of the bump profile while additive does not. Following this analysis, we study the effects of breaking the translation symmetry of the spatially homogeneous network in two ways, using external inputs and using spatially heterogeneous synaptic weights. Effectively, this alters the dynamic landscape of the network from a line attractor to a chain of discrete attractors. External inputs with multiple peaks serve to pin the bump to one of multiple discrete attractors of the network, so that the bump's position evolves as a mean-reverting process. Periodic synaptic heterogeneity also leads to pinning at low modulation frequencies. At high modulation frequencies, the bump can escape from being pinned to a single location in the network, leading to effective diffusion in the limit of long time. We can approximate this effective diffusion using methods for studying a particle diffusing in a periodic potential.

We see the main contribution of this work as introducing the notion of reliability, in the presence of noise, to stationary bumps in neural fields. The specific location of a bump in a neural field carries important information about the stimulus that formed it \cite{amari77,camperi98,laing02}. Noise can degrade this memory, so it is very useful to understand how the architecture and parameters of a neural field model affect how easily this deterioration takes place. This has specific applications in the realm of oculomotor delayed-response tasks in prefrontal cortex, where it is clear there are networks of neurons that can encode visuospatial location during the retention period of such tasks \cite{funahashi89,goldmanrakic95,brody03}. Since our work shows that breaking the translation symmetry of neural fields can serve to decrease noise-induced diffusion of bumps, it is worth pursuing how well this improves the overall memory process. The advantage of a network that is a line attractor is that, in the absence of noise, it can represent a continuum of initial conditions. Since all of these representations are marginally stable, memory is easily degraded when in line attractors when noise is introduced. On the other hand, when symmetry is broken so a network behaves as a chain of discrete attractors, there is a trade-off between initial representation errors versus long term robustness to noise.

Neural fields are known to generate a variety of spatially structured solutions other than bumps, such as traveling waves \cite{wilson73,amari77,benyishai97,bressloff01,coombes12}, stationary periodic patterns \cite{hutt03,curtu04,roxin05}, and spiral waves \cite{huang04,bressloff12}. It would be interesting to study more about how these structures are affected by external noise. It seems that the form of the spatially structured solution markedly contributes to the the way in which noise affects its form and position. Neural fields that support spatially periodic patterns can have the onset of the associated Turing instability shifted by the inclusion of spatially structured noise \cite{hutt08}. In recent work on traveling fronts in stochastic neural fields, it was found that the bifurcation structure of the neural field determined the characteristic scaling of front location variance with time \cite{bressloff12b}. In particular, pulled fronts have subdiffusive variance scaling, as opposed to diffusive variance scaling of a front in a bistable system. We plan to study the effects of noise on bumps in planar neural fields. In this case, the spatial correlations of the noise will be in two dimensions. Therefore, dimensional bias in the synaptic weight or noise correlations could lead to asymmetric diffusion of the bump in the plane. In addition, it is possible this analysis could be extended to two component system, such as a model with local adaptation that generates traveling pulses \cite{pinto01}. If there is a separation of timescales between the activity and adaptation variable, fast-slow analysis might be paired with the small-noise expansion (\ref{rlangzm}) to derive the effective variance in position of the traveling pulse. Finally, it would be quite interesting to study the effects of noise on spiral waves in neural fields \cite{huang04}. Doing so may provide us with some experimentally verifiable measure of whether long-time deviations of the spiral center arise from deterministic meandering or noise.

\section*{Acknowledgements} 
We would like to thank Brent Doiron and Robert Rosenbaum for several helpful conversations concerning this work. ZPK is supported by an NSF Mathematical Sciences Postdoctoral Research Fellowship (DMS-1004422). GBE is supported by an NSF grant (DMS-0817131).

\bibliographystyle{siam}

\end{document}